# High-velocity micro-projectile impact testing

David Veysset,[1,2,a] Jae-Hwang Lee,[3] Mostafa Hassani,[4] Steven E. Kooi,[2] Edwin L. Thomas,[5] and Keith A. Nelson.[6]

[1]Hansen Experimental Physics Laboratory, Stanford University, Stanford, CA 94305, USA;

[2]Institute for Soldier Nanotechnologies, Massachusetts Institute of Technology, 77 Massachusetts Avenue, Cambridge, MA 02139, USA;

[3]Department of Mechanical and Industrial Engineering, University of Massachusetts, Amherst, MA 01003, USA;

[4]Sibley School of Mechanical and Aerospace Engineering, Cornell University, Ithaca, NY 14850, USA;

[5]Department of Materials Science & Engineering, Texas A&M University, 3003 TAMU, College Station, TX 77842, USA;

[6]Department of Chemistry, Massachusetts Institute of Technology, 77 Massachusetts Avenue, Cambridge, MA 02139, USA;

a) Author to whom correspondence should be addressed: dveysset@stanford.edu

ABSTRACT

High-velocity microparticle impacts are relevant to many fields from space exploration to additive manufacturing and can be used to help understand the physical and chemical behaviors of materials under extreme dynamic conditions. Recent advances in experimental techniques for single microparticle impacts have allowed fundamental investigations of dynamical responses of wide-ranging samples including soft materials, nano-composites, and metals, under strain rates up to $10^8$ s$^{-1}$. Here we review experimental methods for high-velocity impacts spanning 15 orders of magnitude in projectile mass and compare method performances. This review aims to present a comprehensive overview of high-velocity microparticle impact techniques to provide a reference for researchers in different materials testing fields and facilitate experimental design in dynamic testing for a wide range of impactor sizes, geometries, and velocities. Next, we review recent studies using the laser-induced particle impact test platform comprising target, projectile, and synergistic target-particle impact response, hence demonstrating the versatility of the method with applications in impact protection and additive manufacturing. We conclude by presenting the future perspectives in the field of high-velocity impact.

## I. INTRODUCTION: FROM MACROSCALE TO MICROSCALE IMPACTS

To gain full comprehensive understanding of a material's behavior, one must test its properties over a wide range of conditions. For engineering applications, materials and systems need to be tested under conditions that span the range expected during practical use, including environmental conditions (atmosphere, temperature, radiation, external perturbation) and operational modes (deformation state and amplitude, cycle, strain rate, fatigue, and life expectancy). Often, a material's properties and performance cannot be established under all realistic operational conditions but have to be inferred from standardized experiments under a limited range of conditions and then extrapolating to more extreme conditions using dimensional analysis and similarity laws (1, 2). In the case of a material's response to deformation, low rate testing can certainly hint at high-speed performance but it has been shown repeatedly that unexpected material behavior can emerge at increasing strain rates and smaller size scales (3–6). Perhaps the first record of such an observation was by Galileo, who noted that a hammer blow leads to effects that would not be observed when the same enormous force is applied slowly under near static conditions (called quasi-static) (7). We intuitively understand that quasi-static testing of an object or a material does not take us far toward an understanding of dynamical behavior. In this review, we will describe experimental impact techniques that have allowed studies of material behavior under dynamic conditions (section II) followed by an emphasis on single high-velocity microparticle investigations using the laser-induced particle impact test (LIPIT) (section III).

Among dynamic testing methods, impact is perhaps the most ancient method and is still widely applied today under much more controlled conditions and expanded capabilities. Historically, the impact engineering field — only recently coined as such in the 1980s — has been driven by the fields of weapon design and tool crafting (8). Impact weapons have evolved hand in hand with protective materials and systems, which therefore necessitate field testing of both the weapon and the armor. Rudimentary weapons have been conceived following the simplistic view that the more massive and rapid the impactor (the larger the translational kinetic energy, KE) the more the potential damage to the target and, likewise, the more massive the protection the better. Weapons are still conceived nowadays in this view, with so-called "kinetic energy" penetrators, consisting of a high-density material core (e.g. tungsten carbide or depleted uranium) for armor-piercing purposes (9, 10). While the medieval wrecking ball is still largely used today, the industrial revolution saw the development of extensive KE impact tools for forging, mining, and construction/destruction applications including drop forging (11) or pile driving (12). In science, the fields of impact physics and shock physics thrived in the mid-twentieth century, not only motivated by a modern and more scientific approach to ballistics and shock but also by more diverse subjects such as meteorite crater formation or spacecraft protection. Most experimental tools for macro-scale high-velocity impacts still in use today were first introduced during this time period.

The most popular and established impact methods for macro-scale testing are undoubtedly pendulum-based impact (Charpy, Izod), drop-weight impact, Taylor impact, ballistic and plate impact tests, in order of increasing achievable strain rates [13]. Pendulum-based Charpy and Izod tests, which differ by their specimen target geometry, and drop-weight tests follow a similar concept where a heavy non-deformable impactor is accelerated by gravity and strikes a target, which is often notched to initiate fracture of the specimen. The methods allow studies and evaluations of impact resistance of materials and composites, mostly plastics and metals, and measurements of fracture toughness, ductility, strength, and energy dissipation [14–18] for a wide variety of loading modes (tension, compression, shear, torsion stress states). ASTM Standards have been written governing the details of experimental testing protocols and the interpretation of results from impacts with different impactor geometry and target materials and configurations (e.g. ASTM E23). Both pendulum-based and drop-weight instruments are widely used in research for material characterization and in industry for material development and quality control. Specimen dimensions typically range from a few millimeters to centimeters. However, impact velocities above a few meters per second are seldom achieved.

Gas-based acceleration techniques can be categorized as a function of projectile and target configuration, each allowing studies of different aspects of impact and shock physics. The Taylor impact test (also referred to as the rod impact test) consists in firing, via a powder or gas driven gun, a mm to cm-sized cylinder, having an aspect ratio L/D of > 5, up to a few kilometers per second toward a massive, non-deformable (highly rigid) target [19]. Because the projectile and not the target is the sample of interest, traditional Taylor testing is considered to be a reverse ballistic technique. Nonetheless, numerous studies also looked at symmetric impacts were both the projectile and the target are rod shaped and of interest [20, 21]. Taylor impacts have mainly been used to establish materials constitutive models. A large range of strain rates can be achieved during testing, with very high strains at the projectile-target interface and reduced strains at the rear-end of the projectile [21–24]. Classical ballistic tests involve spherical, pellet, or bullet-shaped projectiles fired by guns. Upon exiting of the gun barrel, the projectile, usually accelerated in a sabot, detaches with launch velocities up to several kilometers per second, depending on the gas-gun and projectile caliber [25]. Detachment of the projectile from the sabot can be facilitated by rifled barrels that impart spin and associated rotational KE. Studies are largely conducted to assess the ballistic resistance of materials and composites for protective purposes and for the fundamental understanding of material behavior under high-rate deformation and penetration [26–31].

Space exploration is constantly driving experimental and technical development toward higher velocities for μm to cm-sized objects for studies of protection against hypervelocity meteorites and debris, using not only gas-based technology but also electromagnetic systems or a combination of both [32–35]. Plate impact experiments have become the gold standard for testing in the field of shock physics. As in ballistic experiments, plates are accelerated down a gun barrel in a sabot [25]. Traditionally, most plate impact experiments have been performed at normal incidence for easier interpretation and simpler instrumentation; however, increasingly complex configurations are being developed for more complex sample loading with increased experimental control and more extensive diagnostic capabilities [36–38]. Plate impact experiments are of particular interest due to the fact that upon high-velocity impact, shock waves are generated both in the impactor and the target under 1D-strain conditions, until the release of shock waves from the edges of the impactor disturbs the 1D state. Induced pressures can go above hundreds of GPa with associated strain rates above $10^7$–$10^8$ s$^{-1}$. Plate-impact experiments have been used to determine Hugoniot curves of materials (39–41), measure dynamic spall strengths (42, 43), investigate high–pressure phase changes (44) and study shock–induced chemistry (45). Finally, Split-Hopkinson pressure bars are also widely used for shock studies as well as constitutive parameter determination, but compared to the methods mentioned above do not rely on direct impact on the target but rather on shock wave transition through and reflection from an impacted rod and a confined target of interest (46).

Implementing diagnostic techniques complementary to the launch method is key to fruitful and rich experimental investigations, which would otherwise be reduced to merely destroying specimens. Diagnostic methods can be distinguished between contact vs non-contact and in-situ vs ex-situ. Non-contact methods are usually preferred over contact ones to limit specimen response disturbance and instrument failure, and to facilitate sample preparation. For instance, in the case of reverse ballistics as in rod impact experiments, where the sample of interest is the projectile, it is challenging to directly mount contact diagnostics on the projectile (47).

In-situ diagnostics are more desirable than ex-situ diagnostics as impact response is by nature dynamic and post-impact characterizations can hardly reveal the sequence of events. Most non-contact in-situ methods rely on optical techniques such as high-speed photography (including thermal imaging) and interferometry (48), or laser velocimetry (49). These tools can offer high temporal resolution but suffer from poor spatial resolution and are limited to surface interrogation for opaque specimens, although recent progress in femtosecond X-ray measurements has been pushing forward experimental capabilities (50, 51). A comprehensive review on diagnostic methods for macro-scale impacts was written by Field et al. and can be found in Ref (13).

With the rapid development of nano-technologies and the rich emerging micro-mechanisms and mechanical devices at smaller scales, the standard macro-scale projectile launchers are inappropriate because their projectiles are too large. Samples endowed with novel properties through their nano-scale architectures are often produced in small quantities on the order of a picogram with a corresponding volume of ~1 um$^3$ (for a typical polymer density of 1 g/cm$^3$). Considering these dimensions, macro-scale impact experiments are nearly impossible. However, the interesting mechanical responses of these materials often manifest at the microscale through size-dependent behavior and cooperative responses of nanoscale elements (52). To probe such dynamic responses, impact experiments must be conducted at or below the micro-scale. A difference in size scale between the characteristic dimensions of the impactor and the (larger) target also facilitates interpretation.

Micro-scale impacts also prove to be useful in a variety of engineering applications. First of all, hypervelocity micrometeorites and orbital micro-debris represent a threat to the integrity of spacecraft and to astronauts performing extravehicular activities, requiring novel materials design and protection (53, 54). Secondly, the cold-spray additive manufacturing technique uses high-velocity metallic microparticles to build up coatings via impact bonding (55). Thirdly, drug delivery methods using high-velocity drug-loaded microparticles can be used for needle-free epidermal immunization (56). Finally, sand particles carried in pipelines or by the wind can erode inner walls of pipelines (57) or helicopter rotor blades (58). All of these topics call for laboratory experimentation at the appropriate length scale.

While the first experimental techniques in the late 1980s focused on microprojectiles for biological applications (59), the emphasis on nanostructured materials and micro-manufacturing methods motivated the development of novel techniques to address fundamental materials science questions at the nanoscale. Here, we present a review of the techniques for high velocity (up to a few km/s) micro-projectile impact testing, focusing on solid substrate-solid particle impact studies. **Figure 1** shows how complementary experimental techniques cover the field of impact physics to study and develop new applications (from sports and manufacturing to defense and space exploration) while providing insight into material deformation under extreme conditions. While gas-based techniques allow investigations related to ballistics with associated strain rates of the order of $10^3$–$10^6$ s$^{-1}$, Van de Graaff accelerators on the other hand are relevant to hypervelocity impact events encountered in space with extreme strain rates above to $10^{12}$ s$^{-1}$. Micro- and nano-scale launching techniques, inspired by macroscale methods, can be grouped into three main categories: gas-based, laser-based, or electrostatic-based. In particular and as will be elaborated, laser-based methods present the advantage of high throughput and are capable of being operated in small facilities, even on standard laboratory optical tables. Through smaller projectiles and comparable high velocities, these experimental platforms can surpass those on the macro-scale in terms of achievable strain rates, with reported rates up to $10^9$ s$^{-1}$, where distinct behaviors can be observed (60).

In this review, we first describe the widely used methods for microparticle accelerations to high-velocities. We then focus on the laser-induced particle impact test (LIPIT) platform for single-particle impact investigations of materials behavior in the high-velocity regime. We give examples of micro-impact studies conducted with LIPIT focused on target, projectile, and synergistic target-particle impact responses. We conclude with future perspectives in the field of high-velocity microparticle impact physics.

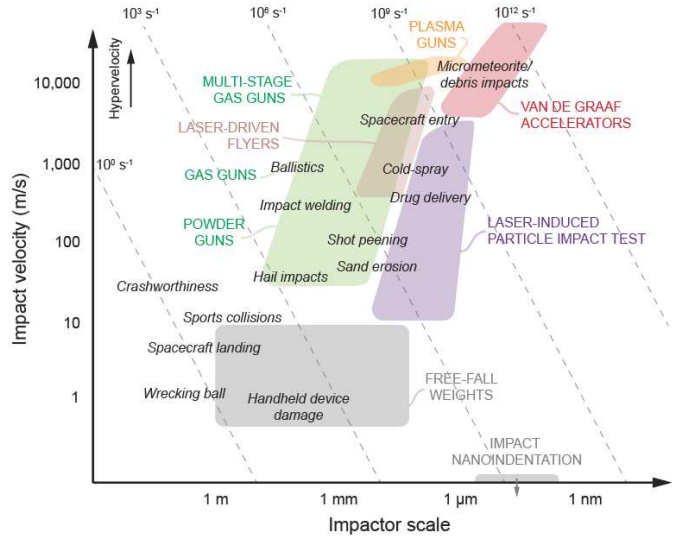

FIG. 1. Impact testing techniques (uppercase) and several applications (italic), along with lines of constant characteristic strain rates. All-capital labels are associated with shaded regions of the same color.

## II. MICRO-SCALE LAUNCHERS

### A. GAS-GUN SYSTEMS

Although gas-gun impact experiments were originally designed for macro-scale projectiles, they have, more recently, been adapted to allow microprojectile acceleration. This technique benefits from a long tradition and expertise in flyer-plate and ballistic impact experiments. Micro-projectile systems require minimum modifications to the gun but attention must be paid to the flyer/sabot design (61). Similar to macro-scale ballistic experiments, these techniques consist in accelerating a sabot (carrying the micro-projectiles) through fast expansion of a compressed gas. In typical two-stage light-gas gun (LGG) systems, a piston is accelerated down a pump tube via fast burning of a powder charge or an explosive detonation. The piston then compresses a light gas, which when sufficiently pressurized, causes rupture of a diaphragm. The light gas expands into the launch tube, which has a smaller diameter than the pump tube, simultaneously accelerating a sabot carrying the micro-projectiles (**Fig. 2a**) (62).

Because the expansion velocity of a gas is inversely proportional to the square root of the gas molecular weight (i.e. the molecular mass), helium or hydrogen are commonly preferred (63). The sabot and microprojectiles separate when the sabot is stopped toward the end of the barrel by a catcher plate (62, 64) or a tapered section, releasing the now detached particles into vacuum at high velocities. Other sabot configurations rely on projectile-sabot separation outside the gun, either through aerodynamic-drag separation of a split sabot or via centrifugal forces acquired through sabot launch in a rifled tube (64).

In most experiments, multiple particles are accelerated in a single shot in so-called buckshot fashion (64, 65). On one hand, running experiments with multiple particles increases hit probability and can generate extensive data acquisition for a single gun fire, however, this is at the cost of impact precision and control. The typical particle velocity spread is of the order a few hundreds of m/s and the aiming angular precision a few degrees. On the other hand, single particle systems are rarer because higher experimental precision is required to limit probability of particle deflection. These systems are nonetheless necessary for applications focused on in single-particle impact events, such as micro-meteorite detection (66). For such purposes, a two-stage LGG was designed at the Japan Aerospace Exploration Agency (JAXA) and spherical aluminum projectiles of 1.0–0.1 mm diameter could be launched at up to 7 km/s (67).

Higher velocities, further into the hypervelocity regime, have been achieved using a so-called three-stage gas gun at the Sandia National Lab (USA) (68). In this system, a sabot is accelerated in a two-stage LGG to impact a stationary flyer target (**Fig. 2b**). The shock generated upon impact travels in the buffer and reaches the flyer that holds the particles. Shock acceleration launches the particles via momentum transfer to yield hypervelocities relevant to micro-meteorite studies (> 10 km/s) (68–70). This system has also been employed to fragment the flyer into micron-sized pieces, with reduced control on projectile size, velocity, and direction (68). Conversely, more compact systems involving compressed gas (with single stage guns) and sabot acceleration have been designed with lower velocity in the range of 100s m/s, with the objective of needle-less drug delivery with drug-loaded penetrating microparticles (62).

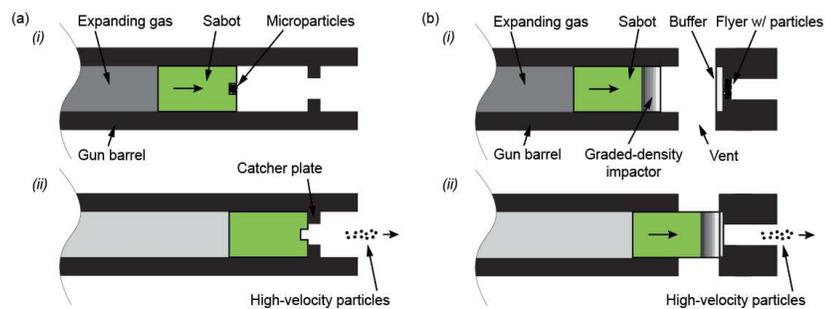

FIG. 2. Gas gun setups for millimeter to micrometer-sized particle launch. (a) *(i)* Typical two-stage LGG: an expanding light gas gun propels a particle-carrying sabot. *(ii)* Upon sabot stoppage, particles separate and continue to travel at high velocities. (b) *(i)* Three-stage LGG: A sabot with a front graded-density impactor, accelerated via light gas expansion, impacts a stationary target (buffer + flyer) holding microparticles *(ii)*. Shock propagation through the buffer to the back surface leads to detachment and acceleration of the microparticles.

In sabot-based acceleration systems, the particle velocity is ultimately limited by the sabot velocity. As long as the particle mass is negligible compared to the sabot mass, sabot speeds are not affected by the particle payload and particles conserve the sabot velocity after separation. Consequently, the particle speed does not rely on the individual particle mass but rather on the sabot acceleration capabilities of the gun (71). However, it is noteworthy that for three-stage LGG systems, the particle velocity is mass-dependent since the acceleration relies on a momentum transfer mechanism (68), which results in even higher particle velocity spread when fragmented flyers are used. Outside of the gun, particle trajectories are affected by environmental conditions. For instance, most gas gun experiments dedicated to material behavior studies are typically operated under vacuum whereas drug-delivery-oriented instruments launch particles under atmospheric conditions. In the latter case, the air drag tends to slow down particles while increasing the velocity distribution.

For most methods described above, particle velocities have been measured either through sabot velocity measurement using photosensors and a time-of-flight approach (65, 72) as for macro-scale impacts or through high-speed photography with nanosecond exposure time. For instance, in the work by Mitchell et al. on microparticle-based drug delivery (71), and related works (73, 74), the particle velocity distribution was measured using particle image velocimetry (PIV). With two ns-duration laser pulses (Nd-YAG) and a single CCD camera, stroboscopic images were taken at the exit nozzle (**Fig. 3a**) and particle velocity distribution could be estimated via image cross-correlation (**Fig. 3b**). Likewise, flash X-ray photography has been used with similar nanosecond time resolution with the added ability to penetrate optically opaque media (68) (**Figs. 3c,d**). Besides particle-velocity measurements, these methods have limited capabilities for real-time in-situ measurements of particle/target interactions and impact responses; studies instead mostly rely on post-mortem examinations (65, 74). Indeed, the regime of high velocity micro-impact, necessitates both micro-scale spatial and nanosecond temporal resolution diagnostics while requiring nanosecond timing and microscale aiming precision for the launch system.

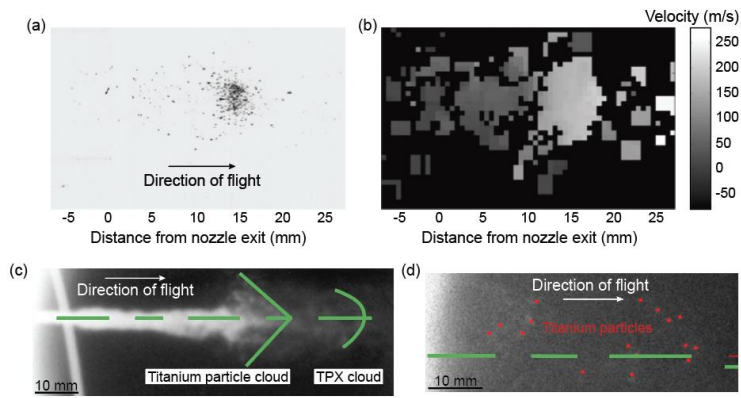

FIG. 3. Double-exposure raw image (a) and corresponding derived PIV map (b) of 99-μm polystyrene particles launched from a single-stage LGG. PIV velocity map yields the instantaneous particle velocity field at the exit nozzle (71). Reprinted with permission from Mitchell *et al.*, Int. J. Impact Eng. **28**, 581 (2003). Copyright 2003 Elsevier. (c-d) X-ray photographs showing 450-μm titanium particles, initially embedded in a polymethylpentene (TPX) matrix, launched using a three-stage gas gun (70). (c) Photograph captured

100 mm from the barrel exit (d) Photograph taken at a later time 440 mm from the barrel exit. The lead particle velocity was calculated to be 8.6 km/s. Reprinted with permission from Thornhill *et al.*, Int. J. Impact Eng. **33**, 799 (2006). Copyright 2006 Elsevier.

## B. SABOT-LESS DRAG-ACCELERATION SYSTEMS

To avoid sabot-related complications (fragmentation, diversion, collection), sabot-less methods have been developed to enable microparticle acceleration. These systems are based on direct interaction between an expanding gas or a plasma jet with the particles, where the drag exerted by the gas or plasma accelerates particles to high velocities. Particles can either be initially positioned in the course of the flow using stationary mounts or directly injected into the flow. For instance, Kendall et al. devised a hand-held contoured shock tube where particles, instead of being mounted on a sabot, were directly enclosed between two diaphragms designed to burst under light-gas pressure (73), as depicted in **Figs. 4a**. They were able to accelerate particles, via drag forces, different materials and diameters (gold 0.2–2.4 µm, polystyrene 11–20.5 µm, glass 2.6 µm) to velocities ranging from 200 to 600 m/s. Motivated by drug delivery applications, particles were intentionally accelerated as a cloud (buckshot mode) to maximize the drug payload by shot (up to 2 mg), inevitably resulting in large velocity distribution (similar to what is shown in **Fig. 3b**). Other works, such as performed by Rinberg et al. (75), have also employed compressed gas but particles were injected in the gas stream though a secondary channel (**Fig. 4b**). Additionally, Rinberg et al. implemented a suction system to recover the helium gas at the exit of the nozzle and to limit gas interaction with the target, which can be harmful when dealing with tissue.

Drag acceleration is also at the center of the cold spray technique, which has been gaining interest in the coating industry since the 1980s and more recently in the additive manufacturing industry (76). A high-temperature compressed gas (typically helium, nitrogen, or air) is used as a propulsive gas to accelerate particles from a powder feedstock, most often toward a metallic target (as shown in **Fig. 4b**). The formation of the deposited coating is via a solid-state process relying on particle kinetic energy (and impact-induced plastic strain and adiabatic heating) rather than thermal energy of preheated particles. Because cold spray industrial systems accelerate simultaneously, in a continuous stream, a large number of particles (**Fig. 5a**) (with a range of particle diameters and impact angles) in a hot (up to T ~ 1000 °C) gas, it has been difficult to answer fundamental physical questions with those systems: The individual particle velocity and temperature are not well-controlled and are mostly deducted from gas dynamics or PIV measurements on many particles. A review detailing velocity measurements in cold spray has been recently written by Yin et al. (77). Owing to their large number, particles can interact multiple times with other particles and the target. It has been therefore difficult to unravel the unit impact-bonding process using these systems. To overcome the complication of particle-particle interactions, wipe tests, where the target is rapidly translated in the plane orthogonal to the spray jet, can be conducted, enabling post-mortem observations and characterizations of single particle deposits (78). However, this technique does not circumvent the lack of information about the history of the particle (particle velocity, impact angle, particle temperature). To solve these complications, other methods for metallic-microparticle impact tests have been implemented (see section 2.3.2).

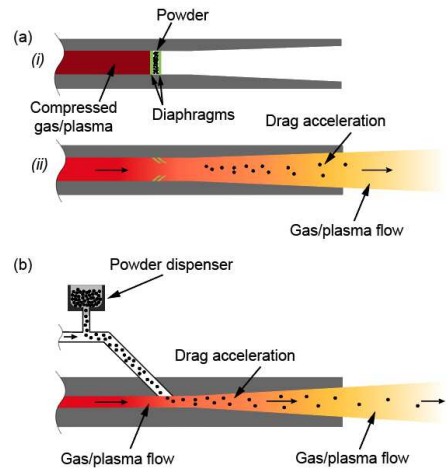

FIG 4. Drag-acceleration systems. (a) A compressed gas ruptures diaphragms holding microparticles, which are subsequently accelerated by the gas flow. (b) Alternative method where particles are injected into the flow.

Following a similar concept, plasma-based systems have enabled purely gas-dynamic acceleration of charge-neutral microparticles (essentially dust) through drag acceleration of a fast-expanding plasma jet with velocities up to a few km/s. The attainable velocities are significantly higher than what is reachable by compressed gas systems (61) and similar to multi-stage gas guns. Coaxial plasma accelerators (guns) were originally developed in the 50s-60s (79) and have been proposed as plasma jet injectors for fusion applications to create, for instance, a spherically-imploding plasma shell (80). Recently, plasma guns have been adapted to accelerate particles as tracers to study plasma dynamics, where plasma properties could be deduced from particle trajectories (81, 82). They have also been suggested as a promising tool for space propulsion or launch (83).

Coaxial plasma guns consist of two concentric cylindrical electrodes. Before the shot, the coaxial gap is filled with a propellant gas (e.g. deuterium), which becomes ionized when a switch connects a capacitor bank to the electrodes. This generates a plasma that is then accelerated by axial Lorentzian forces down the gun, where the highest velocities are achieved near the center rod electrode. Particles can be injected into the plasma inside the bore of the gun, using for instance, a piezoelectric transducer to shake off particles from a dust container into the plasma (83) or directly mounted, statically, in the gun barrel prior to plasma generation (84). Even though relatively high velocities can be reached, the presence of a plasma, which can potentially impact the target and degrade the projectiles, significantly complicates the impact event (**Fig. 5b**).

Gas or plasma drag-entraining methods present the advantages of being able to accelerate large numbers of particles to relatively large velocities (85), allowing material build-up in cold spray or statistical analysis for plasma studies (83). However, they offer limited control of particles' trajectories, owing to gas dynamics instabilities and the presence and impact of hot gas or plasma on targets, which can both heat and chemically alter the target surface. The large uncertainties in impact timing and aiming render these systems inadequate for single-impact observations and systematic investigations of unit impact events. Finally, contrary to sabot-based techniques, the maximum velocities reached through drag acceleration depend on both the mass and the cross-sectional area of the projectiles.

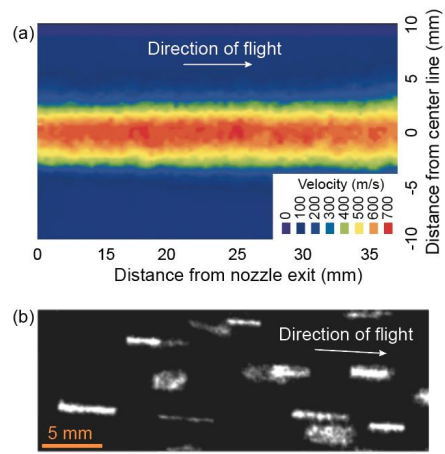

FIG 5. (a) PIV image showing velocity map converted from multi-frame imaging of aluminum particles exiting the helium nozzle of a cold spray system (86). Reprinted with permission from Pattison *et al.*, Surf. Coatings Technol. **202**, 1443 (2008). Copyright 2008 Elsevier. (b) Graphite particles (5–24 μm diameter) accelerated by a coaxial-gun-generated plasma flow. Particles are glowing due to plasma heating. Velocity (0.8–1.5 km/s) can be deduced from elongation of particle due to 12-μs camera exposure time (83). Note that with such long-exposure imaging it is not possible to distinguish single particles. A single streak could originate from two or more juxtaposed particles. Reprinted with permission from Ticoş *et al.*, Phys. Plasmas **15**, 103701 (2008). Copyright 2008 AIP Publishing.

## C. LASER-ABLATION SYSTEMS

In essence, laser-ablation systems rely on the same physics and acceleration mechanisms (sabot acceleration/separation, drag acceleration, momentum transfer) as the gas and plasma guns described earlier. However, focused laser pulses can deliver large energies in much shorter times (down to the femtosecond or possibly less) with high peak power on areas that are much more localized (down to the micro-scale) than conventional methods. These characteristics make them *a priori* well suited for microscopic high-velocity launch, and moreover these approaches offer high controllability, reliability, and safety.

## 1. LASER-LAUNCHED FLYER PLATES

When a laser pulse is focused on an absorbing material with sufficient peak power, the material is flash heated and can liquify, evaporate, sublimate, or be converted to a plasma. The sudden material expansion can serve, similar to gas gun or macro plate impact experiments, as a means to accelerate a flyer. Typically, in a laser-driven flyer (LDF) experiment, a ns-duration laser pulse, e.g., from a commercially available Q-switched Nd:YAG laser with energies up to a few hundred mJ, is focused on a metallic foil (typically few hundred microns in diameter and up to a few tens of microns in thickness) glued (with for example a transparent epoxy) to a thick, rigid substrate that is transparent to the laser wavelength (87–89). Upon arrival of a high-fluence laser pulse, the metal absorbs the laser radiation and a plasma is generated at the interface between the glue and the metallic film. The confined plasma rapidly expands and deforms the metallic foil which subsequently tears off the rigid substrate and is driven off as a flyer plate (see **Fig. 6**), with speeds up to a few km/s (90). Ideal plate experiments require high control on the flyer profile, including its shape and the velocity. In other words, the flyer must remain flat as it travels from the launch pad (LP) to the target. Thus top-hat versus Gaussian focal spots have been preferred for uniform plasma generation and plate acceleration (89). Likewise, short LP-target distances in vacuum are desired to limit plate deformation during propagation due to drag forces (91). In alternative design, metallic films have also been directly coated on optical fibers for their low cost, increased compacity, flexibility, and top-hat beam profile (92).

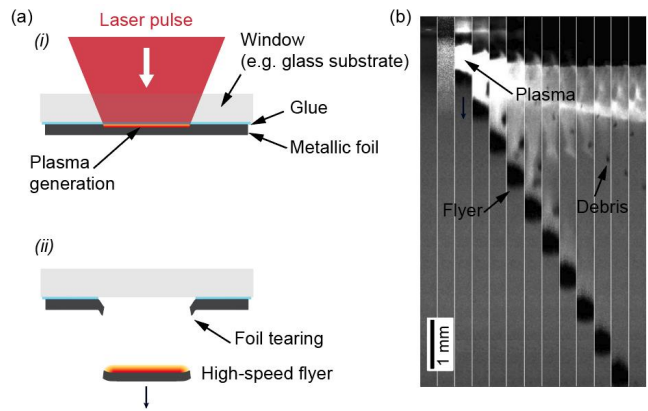

FIG 6. (a) Schematic illustration of LDF launch pad (LP). (i) A laser pulse is focused through a transparent window to the back of a metallic foil, generating a plasma. (ii) The expanding/driving plasma accelerates the foil into free space. Adapted from (91). (b) High-speed photography of a flyer plate launched at 560 m/s. The bright plasma and ablation debris are visible in the images. The interframe time is about 1.6 µs. The exposure (shutter) time is 1 µs, which results in blurring of the flyer (93). Reprinted with permission from Dean *et al.*, Appl. Opt. **56**, B134 (2017). Copyright 2017 The Optical Society.

Earlier investigations, spear-headed by the Los Alamos National Laboratory (LANL) (90) and the Naval Research Laboratory (NRL) (94) in the US, were motivated by interest in impact explosive initiations or fast ignition (90, 95) but recent interest in LDF has recently been growing more broadly in the shock physics community (91, 96, 97). Indeed, LDF experiments are essentially similar to plate impact experiments with considerably shorter shock durations (usually of the order of tens of ns, depending on the plate thickness) but with generated shock pressures that can compete with macro-scale systems (up to ~200 GPa (98)) allowing studies of material behavior under dynamic (shock) loading.

With a relatively simple LP assembly (down to three layers) and micro-scale localization of the laser damage, tens of shots can be performed with a single LP, increasing the number of experiments from a few a day in gas-gun systems to tens of experiments a day for LDF systems. It should also be noted that these experiments can be conducted on a table-top apparatus, in contrast to much larger facilities required by gas-gun setups. On the other hand, inherent challenges remain, particularly regarding the flyer characteristics including its dimensions, planarity, temperature (due to plasma heating) and integrity (extent of material loss through ablation) (96, 99). Experimental efforts have been dedicated to minimize the partial ablation/vaporization of the flyer caused by direct plasma heating and shock-induced spallation. Strategies include the addition of laser shielding layers (e.g. opaque metallic layers) to avoid direct laser exposure and heat shielding layers (using low thermal diffusivity materials, e.g. epoxy). In such multilayer configurations, the flyer is accelerated by the shock generated in the layers underneath it rather than by plasma expansion (96, 100). We note, however, that Curtis et al. have demonstrated that flyer ablation can be limited even in a simple configuration (as depicted in Fig. 6) where the laser energy is mostly absorbed in the glass substrate rather than at the surface of the flyer (89).

Most diagnostic methods are inherited from macro-scale plate experiments, primarily VISAR or photon Doppler velocimeters (PDV) and high-speed photography (89, 93, 101). Increased timing control brought by all-optical systems, allowing high-precision synchronization, has also triggered the development of more elaborate diagnostic capabilities (e.g. in the Dlott group at the University of Illinois at Urbana-Champaign) in shock experiments, including for instance real-time emission spectroscopy (102) and optical pyrometry (103).

While most studies have been centered around the target's shock state, a few works have been devoted to hypervelocity projectile penetration or impact resistance of materials using this technique almost exclusively directed toward protection from space debris (104–106). However, LDF systems have not been the experimental platforms of choice in general for hypervelocity investigations because flat projectiles are rarely encountered in hypervelocity environments.

## 2. LASER-INDUCED PARTICLE IMPACT TEST

Several acceleration mechanisms to achieve fast-moving microparticles had been demonstrated by techniques including laser ablation (LDF in **Fig. 6a**) and laser-induced shock wave (107). However, the first introduction of a laser-driven microprojectile as a precisely-defined high-strain-rate (HSR) microprobe was demonstrated by Lee et al. using the Laser-Induced Particle Impact

Test (LIPIT) (108). As a quantitative HSR characterization method, the key progress advanced by LIPIT is the precise quantification of kinetic parameters of the well-defined microprojectile before, during (*in situ*), and after mechanical interactions with a specimen using ultrafast microscopy with high temporal (nanosecond) and spatial (sub-micrometer) resolution. Although the first demonstration was done using an 800-nm, 250-ps duration ablation pulse from an amplified Ti:Sapphire laser system(108), current LIPIT designs, with setups at MIT (109), Rice University (110), UMass Amherst (111), University of Wisconsin-Madison (112), and NIST (113) use an ablation laser pulse from a Q-switched Nd:YAG system at either the fundamental (1064 nm) or the second harmonic (532 nm) wavelength. The launch pad (LP) assembly includes a glass substrate (typically 200-µm thick), an ablation layer (gold film), and a lightly crosslinked elastomer (polydimethyl siloxane (PDMS) or polyurea, 20–80-µm thick). When the laser pulse ablates the gold film, a plasma is generated and rapidly deforms the elastomer layer, which in turns ejects a particle to high velocity (see **Fig. 7a**). Separation is provided by elastomer retraction by elastic forces as the elastomeric layer remains attached to the LP. With low particle-elastomer adhesion, the particle can freely travel toward a target. The elastomeric layer also serves as an effective thermal barrier between the hot ablation-generated plasma and the projectile, so that the projectile is maintained at its pre-launch temperature.

The first version of the LIPIT initially developed by Lee et al. in 2010 at MIT (108) instead relied on direct drag acceleration, similar to sabot-less systems described in section II.B. A monolayer of silica spheres was deposited on a thin dye-doped polystyrene layer. Upon laser irradiation and absorption by the dye, the polystyrene layer vaporized and multiple particles were accelerated by the unconstrainted expanding gas/plasma (**Fig. 7b**). PDMS launch pads were in use shortly after for much more precise single particle launch and increased velocity control. Aiming for higher velocities, Veysset et al. more recently designed a launch pad eliminating the elastomeric layer, which partially dissipates the plasma-induced acceleration forces (114). Higher velocities could be obtained for heat-resistant ceramic particles via direct plasma drag acceleration. However, metallic particles (Al and Sn) melted due to the direct contact with the hot plasma, which prevented acceleration to supersonic velocities. In the emergent LIPIT technique, the primary strategy is high speed photography either via single-detector multiple exposure with pico/femtosecond laser pulses (**Fig. 7b**) (108, 110, 115) or using a high-speed multi-frame camera (**Fig. 7c**) (109, 116).

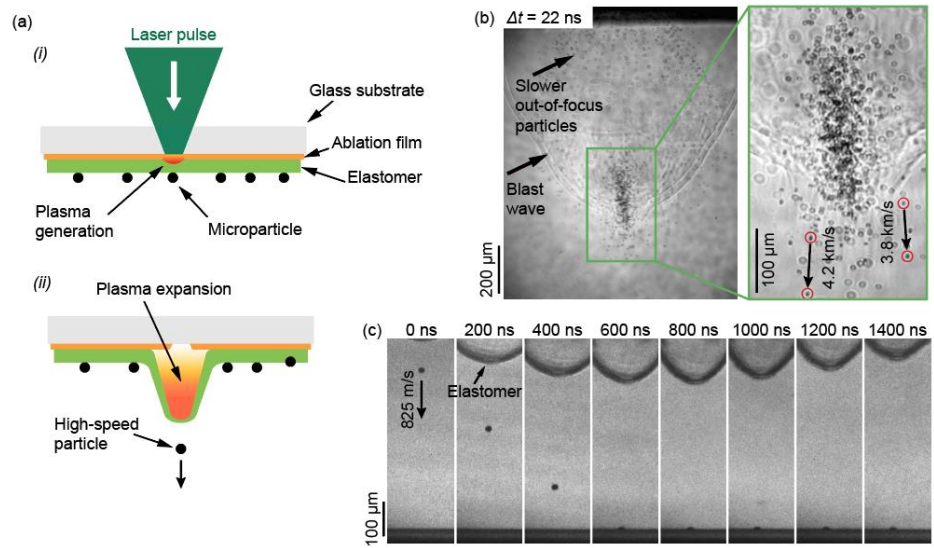

FIG 7. (a) Laser-induced particle impact test (LIPIT) schematic illustration adapted from (109). (i) A laser pulse is focused on an ablation film that is quickly converted to a plasma. (ii) The plasma expands and pushes the elastomeric layer, which accelerates a selected particle to high velocity. Adapted from Veysset *et al.*, Sci. Rep. **6**, 25577 (2016) under the terms of the Creative Commons Attribution 4.0 International License. (b) Double-exposure image showing acceleration of multiple particles (3.7-µm diameter silica spheres) to a few km/s. With a known time separation (22 ns) between the two illumination flashes (250-fs duration laser pulses) the speed can be calculated by tracking particle displacement (108). Reprinted with permission from Lee *et al.*, Nat. Comm. **3**, 1164 (2012). Copyright 2012 Springer Nature. (c) Multi-frame image sequence showing an aluminum particle (12-µm diameter) as it impacts an aluminum substrate with a speed of 825 m/s, subsequently adhering to the target surface. Timing is shown at the top of each frame. The expanding and retracting polyurea elastomer is also visible in the images (116). Reprinted with permission from Hassani-Gangaraj *et al.*, Scr. Mater. **145**, 9 (2018). Copyright 2018 Elsevier.

With LIPIT, particle velocities can be tuned by adjusting the laser energy. The highest velocity reported is about 4 km/s with silica particles (3.7-µm diameter) (108). Contrary to sabot-based gun schemes, the mass of the particle is not negligible compared the mass of elastomeric layer in LIPIT, which is why maximum particle velocities depend on the particle mass (see section II.F.). It should also be noted that with LIPIT, particles can also be accelerated to relatively low velocity down to less than 1 m/s (unpublished), given sufficiently low adhesion between the particle and the polymer. In contrast, because the LDF relies on foil ablation, the plasma must drive the foil sufficiently to detach the foil into a flyer, which results in a higher minimum velocity (>0.1 km/s) (91).

LIPIT has emerged as a fruitful tool for high-velocity micro-impact studies and has been applied for to the study of a wide range of material behavior—for both projectile and target—, in bulk polymers, gels, and metals as well as nano-composites, 2D layers, and fibers. Representative examples of LIPIT studies are presented in more detail in section III.

## D. ELECTROSTATIC ACCELERATION SYSTEMS

The use of an electrostatic acceleration method to launch microparticles to hypervelocity was first described by Shelton et al. in 1960 (117). The method was originally developed to study hypervelocity impacts of single microparticles and has been used extensively and principally in the field of micrometeorite impact studies (118–120) and impact-induced plasma physics (121). Due to the high voltages required for both particle charging and acceleration, this acceleration technique must take place in vacuum, which not only provides a space-like environment but also frees the particle launching process from any concurrent blast wave or propellant debris that must be accounted for in most other techniques. Electrostatic acceleration is achieved by first charging dust or microparticles and then passing the charged particles through an electric potential difference. The ultimate velocity ($v$) that can be reached for a given particle is based on its charge-to-mass ratio ($m/q$) and the magnitude of the potential difference ($U$), where the kinetic energy of the particle $1/2\, mv^2 = qU$ (117). Therefore, it is desirable to obtain as high of a charge-to-mass ratio as possible. The most efficient and effective route to reaching high charge states is to bring a microparticle into contact with an electrode held at high potential. The charging limit for most materials is the threshold for ion evaporation (for positively charge particles) or electron field emission (for negatively charged particles). This also brings up a key limitation to this acceleration technique, namely that at least the surface of the particle needs to be conductive. Metallic particles are often used as well as conducting-polymer-coated (122) or metal-coated insulating particles (123).

There is a wide variety of electrostatic acceleration instruments reported in the literature over the years. They all include a dust or particle source of single highly charged particles for acceleration and impact studies. Examples of dust/particle source designs can be found in refs (120, 124). The charging of particles depends on the particle characteristic and dimensions (surface area), which results in different empirical mass-velocity relationships for different particle materials and dust sources (as illustrated in **Fig. 9**, red solid lines). Often, many more charged particles are generated in these sources than make it into the acceleration region of the instruments. A series of pinhole apertures and electrostatic lenses are used to exclude any particles that are not moving along the axis of the instrument. After a particle is extracted from a dust source, it is accelerated by passing through a large potential difference. In instruments that have demonstrated the highest velocities, the voltage necessary is provided by a Van de Graaff generator (124), as illustrated in **Fig. 8a**. These are typically found in large shared-use facilities or national labs and require a large amount of laboratory space, extremely high voltages (up to 2–3 MV), and specialized equipment (35, 125–127).

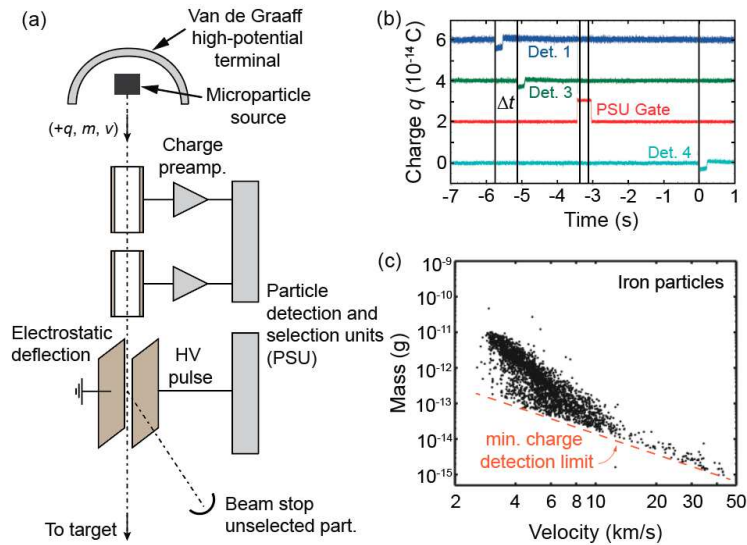

FIG 8. (a) Schematic illustration of a Van de Graaff (VdG) microparticle accelerator adapted from (125). A large potential difference between the high-potential VdG terminal and a ground terminal accelerates charged particles toward a target. Hollow metal cylinders detect particle charge to allow velocity measurements and mass estimations. Unwanted particles can be deflected further down the beam path before reaching the target. Other deflection setups deflect only selected particles toward the target. Reprinted with permission from Keaton *et al.*, Int. J. Impact Eng. **10**, 295 (1990). Copyright 1990 Elsevier. (b) Example of a single-particle signal generated upon passage through multiple charge detectors and PSU gate selection(120). Reprinted with permission from Shu *et al.*, Rev. Sci. Instrum. **83**, 581 (2003). Copyright 2003 AIP Publishing. (c) Mass-velocity distribution obtained using a 2-MV VdG accelerator for iron particles (128). Particle with low charge fall below the detector noise level and are not detected. Reprinted with permission from Lee *et al.*, Int. J. Impact Eng. **44**, 40 (2012). Copyright 2012 Elsevier.

The use of highly charged particles in the electrostatic acceleration technique allows various types of charged particle detectors to be used to measure charge and velocity and to extract particle mass. A series of image charge detectors are often employed to determine particle properties after acceleration and before impact (120) (**Fig. 8a**). An example of a pick-up signal is shown in **Fig. 8b**, where the magnitude of the signal reveals the particle charge $q$ and the delay between detection times $\Delta t$ gives the velocity. After detection and characterization of the particle parameters, electrostatic deflection systems can be employed to allow only particles of a selected charge and velocity, and therefore mass, to impact a chosen target (**Fig. 8a**). Particles that do not meet the desired parameters can be deflected off the axis of the experiment. This is a critical capability due to the fact that most of these instruments launch a stream of particles and post-impact analysis would be difficult without particle discrimination and characterization. In principle, single-particle impact experiments can be conducted using Van de Graaff accelerators (VdGA). However, since particles are ejected from the dust generator at random times, in-flight detection of flying particles is necessary to trigger subsequent data acquisition, limiting real-time diagnostic capabilities. Single-particle discrimination following in-flight detection requires a low rate of particle entry into the acceleration column and hence limits the experimental rate to a few experiments a day. A higher impact rate with lower particle selectivity (wider selected mass and velocity distribution) has often been preferred (120, 124).

Of all of the particle acceleration techniques described in this review, electrostatic acceleration provides the highest achievable velocities, up to several tens of km/s (see **Fig. 8c**) and delivering impacts at characteristic strain rates of $10^{12}$ $s^{-1}$ or higher (129). The technique is constrained by the requirement of using conductive particles and is typically limited to relatively small particle sizes of under a few microns in diameter.

### E. OTHER METHODS FOR PROJECTILE ACCELERATION

The set of methods described in the previous sections is certainly not exhaustive but includes the most widely adopted techniques for microparticle launch. Other methods have been developed motivated by the desire to reach even higher speeds to 1) replicate hypervelocity impacts, 2) develop a space launch platform of cm-scale objects, and 3) create higher shock pressure conditions by hypervelocity impact. Because these methods aim for hypervelocities (beyond the high velocity studies presented here), we do not describe them in detail. Some of these methods, however, are worth of notice. For instance, the railgun has long been seen as a promising technique for hypervelocity launch (130). It involves accelerating a piston-like armature using Lorentzian forces, as for a plasma gun. Velocities up to 5–7 km/s for projectiles with masses of the order of one gram have been demonstrated, but this technique has not surpassed gas guns' more consistent performance until now (131). The inhibited shaped-charge launcher (SCL) and other explosive-based techniques have long been used to study hypervelocity impacts (132). A SCL consists in the collapse of a shock-driven liner to create a micrometric jet (cylinder-like shape), with very limited control on shape, mass, velocity, etc. A brief review, written by Schneider and Schäfer, on these hypervelocity techniques can be found in Ref (35). The Z-Machine, also using Lorentzian forces, have enabled the acceleration of micrometer-thick plates to hypervelocities above 10 km/s for equation of state measurements at pressures reaching ~1000 GPa (133).

### F. LAUNCHER PERFORMANCE SUMMARY

A representative sample of maximum reported velocities as a function of projectile mass for different impact-testing systems over the past 25 years is presented in **Fig. 9**. Launch data are also categorized based on projectile shape and shot mode. Single-shot mode refers to cases where single-particle impact signatures can be recovered whereas buckshot/continuous mode represents cases where single-particle discrimination is not possible or reliable. We note here again that sabot-based techniques such as single-stage and two-stage techniques accelerate particles almost independently of their mass whereas for drag-based (plasma- and gas-drag accelerators, cold spray) and momentum-based techniques (3-stage LGG); lighter particles reach higher speeds. Van de Graaff accelerators, LDF, and LIPIT methods demonstrate a stronger dependence on particle mass which is related to the maximum particle velocity through a nearly constant kinetic energy as shown in Fig. 9. In the high-speed microprojectile regime (shown with a dashed square), LIPIT is uniquely positioned for single-projectile studies.

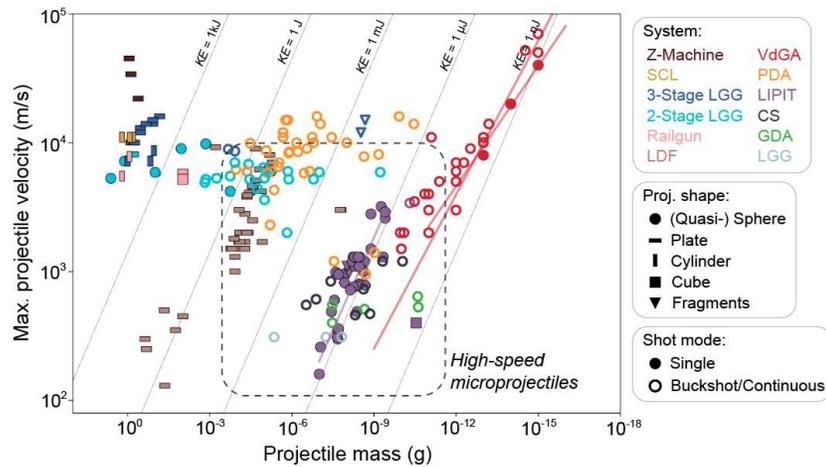

FIG 9. (a) Launcher performance summary, including Van de Graaff accelerator (VdGA) and reported empirical mass-velocity laws for iron and conduction-polymer-coated latex particles (red solid lines), plasma drag and gas drag accelerators (PDA, GDA), laser-induced particle impact test (LIPIT) and empirical mass-velocity law (violet solid line), cold-spray systems (CS), single- and multi-stage light-gas guns (LGG), laser-driven flyers (LDF), inhibited shaped-charge launchers (SCL), railguns, and the Z-Machine. Systems are identifiable by symbol color. Symbol shapes represent projectile shapes. Open-symbols are for buckshot and quasi-continuous feed (or buckshot) systems. Solid symbols are for single-projectile systems. Lines of constant kinetic energy (KE) are plotted. See supplementary material for full data set and references.

## III. HIGH-VELOCITY MICRO-IMPACT BEHAVIOR OF MATERIALS

Here we describe and look into selected material behaviors that were revealed through high-velocity micro-impacts. We focus on the high-velocity regime and not on the hypervelocity regime (> 10 km/s) where the behavior is more extreme and where materials typically exhibit hydrodynamic behavior or are converted to plasma. The intermediate regime of high velocity has recently gained interest with the active and rapid development of LIPIT platforms which are uniquely positioned in this regime (see **Fig. 9**), allowing materials study under previously under-explored deformation regimes. In the following sections, we distinguish three idealized impact configurations where deformations are (i) predominantly in the target (hard particle on compliant target), (ii) predominantly in the particle (compliant particle on hard substrate), and (iii) partitioned between particle and target, which have similar mechanical properties.

### A.  TARGET RESPONSE

## 1. DYNAMIC HARDNESS OF BULK MATERIALS

Similar to what is routinely done in macro-scale ballistic impact experiments, hard/non-deformable projectiles can be aimed at 3D quasi-semi-infinite targets. LIPIT was initially developed to study penetration resistance of a semi-infinite nano-layered material (see next section) and soon adapted to bulk, isotropic materials. Following the premise of increased dissipation through dynamic glass transition of elastomer by Bogoslovov et al. (134), several works have been dedicated to studying the impact and high-rate behavior of elastomers, namely polyureas, polyurethane, and poly(urea-urethanes (PUUs)), under LIPIT to identify and leverage molecular attributes for energy dissipation. For instance, Veysset et al. performed a series of experiments to visualize microparticles impacts on PUUs (135, 136). High-velocity impacts resulted in particle penetration and subsequent rebound with no permanent damage (see **Fig. 10a**). They demonstrated a correlation between the hard segment content of the PUUs and both the coefficient of restitution (ratio of impact velocity and rebound velocity, indicator of energy dissipation) and the maximum penetration depth, related to material hardness. Further works by Hsieh et al., Wu et al., and Sun et al. aimed to elucidate the dynamic stiffening characteristics of PUUs in view of segmental relaxation dynamics to unravel the role of hydrogen bonding in the dynamic glass transition and impact response (137–139). In particular, the slower relaxation dynamics of the soft phase, undergoing a change from rubbery behavior at ambient conditions to a leather/glassy behavior under high rate deformation, was suggested to act as preferential molecular pathways for energy dissipation. Recently, temperature-controlled experiments by Sun et al. on polyurea revealed an increased energy dissipation (and related hardness) at the dynamic glass transition temperature (see **Fig. 10b**) (140). The molecular relaxation at the origin of dissipation was identified as the $\alpha_2$ relaxation corresponding to the soft segments near the hard-segment interfaces, as identified via broadband spectroscopy. This study therefore quantitatively confirmed the macro-scale experiments and thesis by Bogoslovov et al (134). Other soft materials beyond polymers, such as gels or biomaterials, have also been tested with LIPIT to determine their rate-dependent properties. Notably gelatin and synthetic gels, including poly(styrene-b-ethylene-co-butylene-b-ethylene) (SEBS) polymer-based gels with non-aqueous solvent, with varying concentrations were investigated (141, 142). Particle trajectories were observed in the transparent specimen (**Fig. 10c**) and particle trajectory models were evaluated. Recent experiments by Veysset et al. allowed the development of a refined model for particle impact in yield-stress fluids at intermediate Reynolds number (**Fig. 10d**) (142). In all gels tested, it was shown that, under micro-scale testing, the resistance to penetration, related to materials strength, significantly increased compared to macroscale experiments by orders of magnitude, evidencing rate strengthening effects. Further insights into gel response can be achieved by numerical simulations where models are calibrated using particle trajectories and cavity dynamics (143).

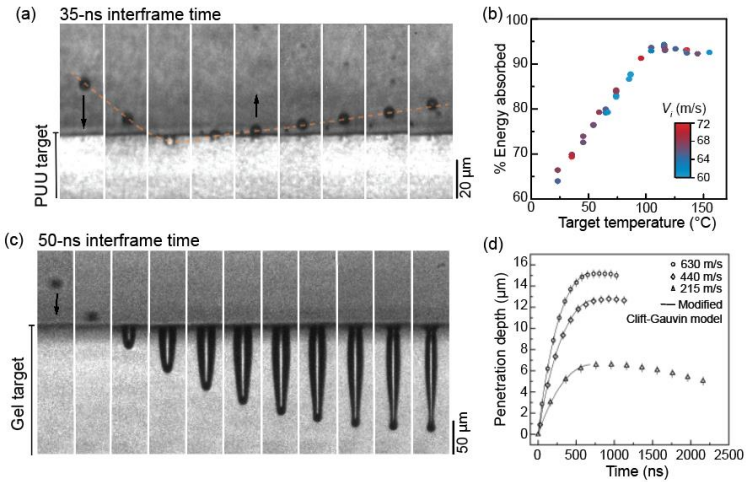

FIG. 10. (a) Multi-frame impact sequence showing a silica particle (7.4-μm diameter) impacting a poly(urea urethane) target at 770 m/s and rebounding at reduced velocity (109). Reprinted with permission from Veysset *et al.*, Sci Rep. **6**, 25577 (2016) under the terms of the Creative Commons Attribution 4.0 International License. (b) Percent impact energy absorption of polyurea target as a function of temperature, revealing glass-to-rubber transition around 115°C (140). Reprinted with permission from Sun *et al.*, Appl. Phys. Lett. **117**, 021905 (2020). Copyright 2020 AIP Publishing. (c) Multi-frame impact sequence showing a steel particle (13-μm diameter) penetrating a SEBS-based gel target at 630 m/s (142). (c) Penetration trajectories in gel for three impact velocities and corresponding modeled trajectory curves. Reprinted with permission from Veysset *et al.*, Exp. Mech. **60**, 1179 (2020). Copyright 2020 Springer Nature.

The dynamic impact hardness of metals has mainly been measured over the past two decades through impact indentation using pendulum-based dynamic micro-indenters (PDMI) (144). PDMI has allowed fundamental studies on materials properties, despite experimental calibration complications and debate over hardness calculations procedure (141, 145, 146). For example, Trelewicz and Schuh demonstrated the breakdown of the Hall-Petch strength scaling in nanocrystalline Ni–W alloys and subsequent emergence of an inverse Hall-Petch weakening regime (147). However, in PDMI, the strain rate is limited to about $10^4$ s$^{-1}$ because of relatively low impact velocities (up to a few mm/s) (145). In contrast, with higher achievable velocities, high-rate impact indentation studies can be conducted with LIPIT as demonstrated by Hassani et al. (148). In that study, hard/non-deformable alumina particles were used to impact pure copper and iron targets at velocities up to 800 m/s. Real-time imaging helped determine the plastic work leading to post-impact imprints on the sample surface (**Fig. 11a**), whose volumes were measured by confocal microscopy (**Fig. 11b**). Using the conventional definition of hardness used in PDMI experiments, defined as the ratio of plastic work to indentation volume, hardness values could be estimated for characteristic strain rates of the order of $10^6$–$10^7$ s$^{-1}$ (**Fig. 11c**). The material hardness was shown to significantly increase with strain rate above $10^3$–$10^4$ s$^{-1}$, which was attributed to a transition in the deformation mechanism from thermally-activated to drag-dominated dislocation motion.

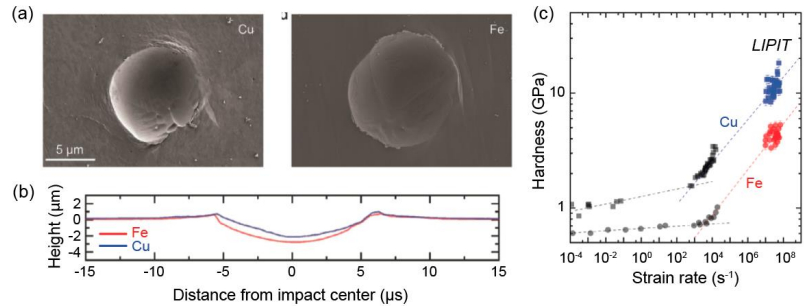

FIG 11. Dynamic impact hardness determination. (a) Post-impact craters and (b) surface after alumina-particle impacts on a copper substrate at 425 m/s and an iron substrate at 657 m/s. (c) High-rate dynamic hardness as a function of deformation rate (148). Reprinted with permission from Hassani *et al.*, Scr. Mater. **177**, 198 (2020). Copyright 2020 Elsevier.

## 2. HIGH STRAIN RATE RESPONSES OF NANO-SCALE MATERIALS

Materials exhibiting nanoscale size effects or nanoscale phases are envisioned to be a breakthrough toward achieving high energy absorption for lightweight protective materials due to their mechanical anisotropy and high specific inter-phase interaction. As stated earlier, LIPIT was first devised and applied for detailed observations of polymer layered nanocomposites subjected to supersonic micro-projectile impacts (108). The HSR responses and associated energy dissipation mechanisms of studied nano-scale materials have been categorized into two cases: primarily localized and primarily delocalized responses as illustrated in **Fig. 12**. Although the actual responses of the materials generally include both responses, this classification based on materials' dominant deformation scale is useful to understanding energy dissipation mechanisms depending on strain rates, as well as material properties. Thus, the penetration energy ($E_p$) can be expressed by $E_{plug} + E_l + E_d$, where $E_l$ is the localized energy dissipation at a vicinity of the direct impact area and $E_d$ represents delocalized energy dissipation beyond the direct impact area. The approximated kinetic energy transferred to a plug (as depicted in **Fig.12**), $E_{plug}$, is supposed to equal $\rho h A_p v_{plug}^2/2$, where $\rho$, $h$, and $v_{plug}$ are the specimen's mass density, thickness, and a velocity of a plug, respectively. The direct kinetic energy transfer to a plug of a specimen has been observed in penetration of free-standing thin-films of polystyrene (3, 149), polycarbonate (150), multilayer graphene (151, 152), and graphene nanocomposites(153) with hard, solid silica projectiles. A material's physical responses become localized near the projectile when (i) the projectile speed is fast or comparable to the propagation speed of deformation ($\propto \sqrt{E/\rho}$ or $\sqrt{G/\rho}$) and (ii) the onset of material failure at the direct impact area (or projectile's cross-sectional area $A_p$) is faster.

It was shown that for polymers having relatively slow propagation speeds of deformation $E_p$ can be effectively increased by enhancing $E_l$. A higher entanglement density and a higher molecular weight of polymers can delay the onset of brittle fracture at the direct impact area (149) and localized adiabatic heating can create a high rate visco-plastic flow around the perimeter. Thus, further plastic deformation around the periphery of the projectile increases $E_l$ (**Fig. 12a**) (3, 150). In case of elastic materials having high tensile strength and a low density, $E_p$ can be increased via enhancing $E_d$ and localized plastic deformation is relatively insignificant. For example, multilayer graphene (MLG), having an exceptionally fast propagation speed of deformation (tensile wave speed of 22 km/s), can rapidly transfer the projectile's kinetic energy to neighboring areas through the impact-induced conic deformation (**Fig. 12b**) (151). Although this energy transfer is not energy dissipation through heat but rather, energy delocalization, the transferred energy has been suggested to eventually dissipate through various secondary processes such as cracking, folding, crease, and aerodynamic friction.

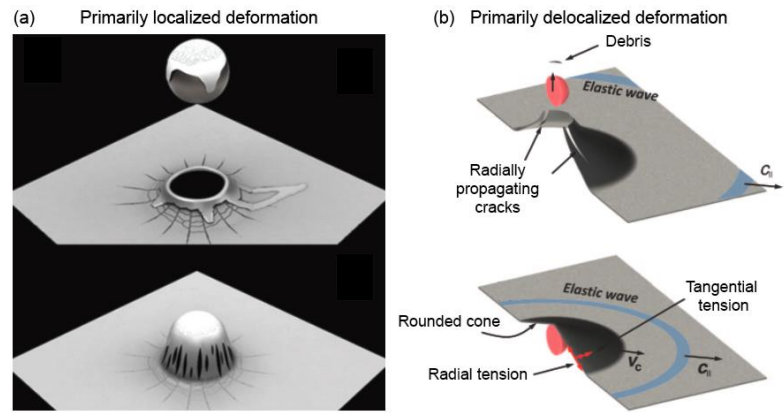

FIG 12. Schematic illustration of energy dissipation mechanisms during projectile perforation of materials primarily through localized (a) (3) and delocalized response (b) (151). (a) Reprinted with permission from Hyon *et al.*, Mater. Today **21**, 817 (2018). Copyright 2018 Elsevier. (b) Reprinted with permission from Lee *et al.*, Science **346**, 1092 (2014). Copyright 2014 The American Association for the Advancement of Science.

Due to the high temporal (sub 100 ns) and spatial resolutions (sub 100 nm) required for the quantitative analysis of deformation in LIPIT studies of nano-scale materials, the real-time characterization of deformation has been challenging. However, because post-impact damage features at the vicinity of an impact or penetration region also provide essential information about characteristic deformation modes and energy dissipation mechanisms, SEM studies of post-LIPIT specimens are generally conducted. **Figure 13** catalogs several representative damage features originating from relevant energy dissipation mechanisms. In a glassy polymer system such as polystyrene (PS), because $E_p$ is increased by higher fracture toughness, higher molecular weight, it was shown that PS can dissipate more energy via more extended radial crazes (**Fig. 13a**). In contrast, polycarbonate (PC) has significantly higher entanglement density than PS does, yielding a primary energy dissipation mechanism of PC, which can be seen from craze-less plastic deformation in **Fig. 13b**. For an anisotropic material with long-range ordering, the deformation responses are greatly affected by the angle of impact direction relative to the structural orientation.

For example, a bulk lamellar nanocomposite consisting of glassy and rubbery layers dissipates impact energy via strong reorientation events when the impact direction is parallel to the layers, while extreme lamellar compression is a dominant energy dissipation mechanism for the perpendicular orientation (**Fig. 13c**). Moreover, the nanocomposite of the perpendicular orientation demonstrated 30% shorter penetration depth compared to the perpendicularly oriented case, indicating higher energy dissipation performance. Compared to the polymers, high strength elastic nanomaterials such as MLG do not show considerable plastic deformation features but relatively small number of straight radial cracks (typically less than 6), propagating a long distance, multiple times the projectile's diameter (**Fig. 13d**). The long cracks and large penetration opening imply that extensive conic deformation was made via impact energy delocalization (**Fig. 13b**). Because the energy delocalization is achieved via prompt conical deformation, the presence of the surrounding fluid (i.e. air) restricts the conic deformation and reduces $E_d$. Thus, MLG under vacuum can increase its energy delocalization performance to be 3 times higher than in air (152). Despite the outstanding performance of MLG in enhancing $E_d$, due to its elastic and crystalline nature, MLG has the typical weakness of crystalline ceramics: low fracture toughness and high susceptibility to local defects. As a result, the onset of structural failure at the impact region is sensitive to local defects and this local failure progresses to a global failure through rapidly propagating cracks. However, the weakness of MLG can be noticeably decreased by quasi-plastic deformation. As seen in **Fig. 13d**, when the initiation of cracks at the impact region is suppressed by quasi-plastic deformation through localized folding, interlayer sliding, and wrinkling of delaminated layers of MLG, improved delocalization is possible.(152) Furthermore, one can envision the maximization of $E_p$ via the combination of local plasticity and global elasticity. Indeed, nanocomposites of plastic polymers and elastic nanomaterials are suggested to accomplish the simultaneous enhancement of $E_l$ and $E_d$.(153) The nano-scale laminates comprised of alternating layers of silk fibroin and graphene oxide flakes exhibit enhanced hybridized dynamic responses when graphene oxide flakes reach their percolation threshold over which impact stresses can transmit through the continuous phase of graphene oxide flakes. Moreover, the fracture toughness at the impact region is also improved as observed from a remarkably smaller mass loss (**Fig. 13e**).

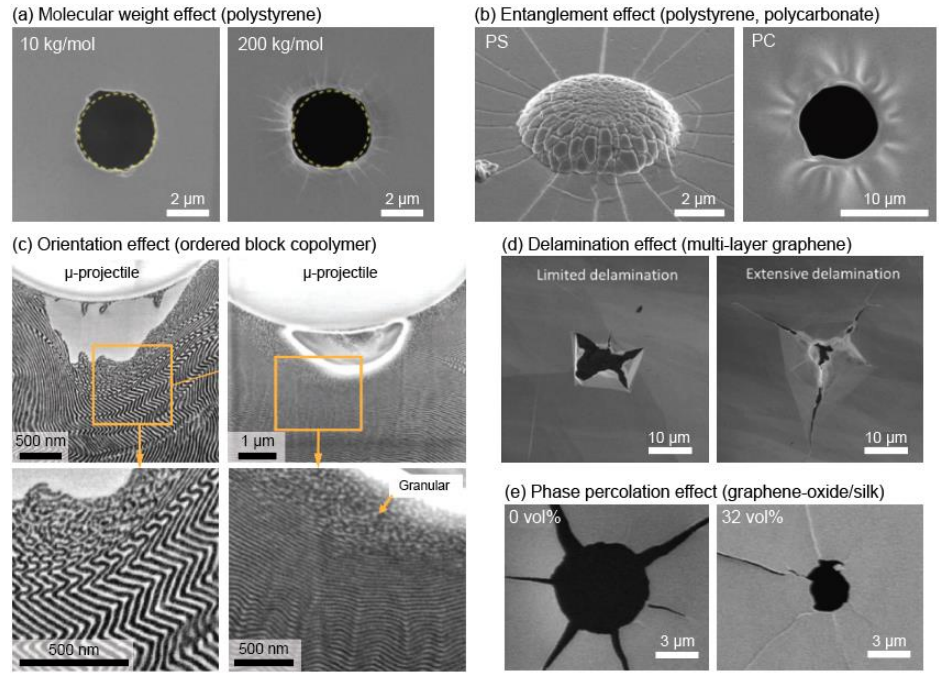

FIG 13. Post-impact damage and deformation features. of various materials. (a) Dense radial and tangential crazes appear around the penetration hole of a high molecular weight polystyrene (PS) freestanding membrane (149). Reprinted with permission from Xie et al., Macromolecules. **53**, 1701 (2020). Copyright 2020 American Chemical Society. (b) Because of higher entanglement density, polycarbonate (PC) shows yield-dominant deformation compared to the craze-dominant features of PS (3). Reprinted with permission from Hyon et al., Mater. Today **21**, 817 (2018). Copyright 2018 Elsevier. (c) Deformation features of block copolymer PS-b-PDMS lamellae when the impact direction is parallel (left) or perpendicular (right) to the plane of the lamellae (108). Reprinted with permission from Lee et al., Nat. Comm. **3**, 1164 (2012). Copyright 2012 Springer Nature. (d) Higher energy dissipation of MLG is observed when delamination occurs in the penetration process (152). Reprinted with permission from Lee et al., Science **346**, 1092 (2014). Copyright 2014 The American Association for the Advancement of Science. (e) When graphene-oxide flakes are added over the percolation threshold (32 vol%), smaller penetration damage and larger energy dissipation are possible (153). Reprinted with permission from Xie et al., Nano Lett. **18**, 987 (2018). Copyright 2018 American Chemical Society.

Besides the membrane-like (2D) and the semi-infinite (3D) specimens, fiber-like (1D) specimens has also been characterized by LIPIT. Compared to 2D geometries, because precise measurement of instantaneous positions of a projectile is possible via ultrafast stroboscopic imaging while interacting with a specimen, energy dissipation and force exchange can be quantified by first and second derivatives of the time-dependent positions (**Fig. 14**). Using LIPIT, the collective dynamics of a CNT fiber, an ensemble of weakly interacting, aligned CNTs, were directly compared by Xie et al. to nylon, Kevlar, and aluminum monofilament fibers under the same supersonic impact conditions (115). Although individual CNTs are an elastic strain-rate-insensitive material, strain-rate-induced strengthening arising from interactions between the individual carbon nanotubes was observed. Moreover, as the CNT fiber surpassed the other three conventional fibers including Kevlar in specific energy absorption under the equal conditions, the demonstrated performance of the CNT fiber at the microscopic scale can also be realized in the scaled-up conditions.

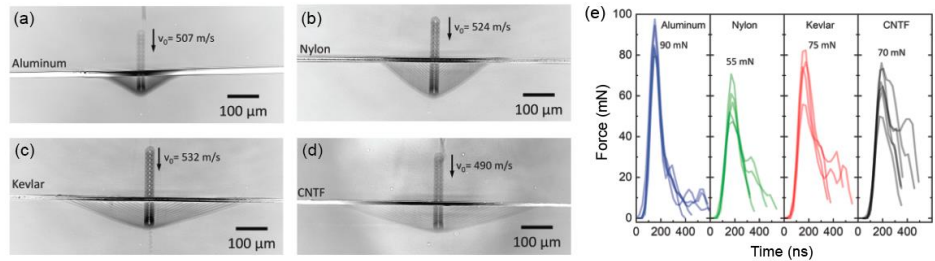

FIG 14. In-situ measurement of force exchange between a micro-projectile and a fiber specimen. Ultrafast stroboscopic micrographs of (a) pure aluminum, (b) nylon 6,6, (c) Kevlar KM2, and (d) CNT fibers under microsphere impact around 500 m/s. (f) Instantaneous forces between the microspheres and the fibers are measured by tracking the positions of the projectiles (115). Reprinted with permission from Xie *et al.*, Nano Lett. **19**, 3519 (2019). Copyright 2019 American Chemical Society.

## B. PROJECTILE RESPONSE

In a reversed ballistic configuration, the dynamic yield strength of a projectile can be determined by analyzing its residual geometry after an impact onto a rigid flat target. This is the basis of the experiment conducted by Taylor (19) using cylindrical projectiles during World War II. Taylor impact test produces a non-uniform plastic deformation with the front part of the projectile crumpling upon impact and the rear part remaining undeformed. Considering the propagation of the elastic and the plastic waves into the projectile Taylor explained the residual deformation and demonstrated that the shortening of the cylinder is proportional to the square of the impact velocity and the inverse of the dynamic yield strength. The original development was subsequently extended by other researchers (154–160), making Taylor impact test one of the main experimental approaches to determine the constitutive behavior of materials at high strain rates. While classically conducted at macroscales, Taylor impact test has inspired LIPIT researchers to study mechanics and materials phenomena with Taylor-like impact experiments at microscales. The decrease in the projectile size from mm in the classic Taylor impact test to μm in LIPIT while maintaining the same level of impact velocities comes with the benefit of extending the achievable strain rates, form nominally ~$10^5$ in the former to ~$10^8$ s$^{-1}$ in the latter.

**Figure 15a**, from Xie et al. (111), shows the flattening of five initially spherical polycrystalline Al 6061 microparticles as a result of impact onto a rigid sapphire target at various velocities. Post-impact characterization of the microstructures revealed localized deformation at the bottom of the particles along with insignificant plastic deformation on the top. In the presence of such non-uniform plasticity, the analytical determination of material's constitutive parameters requires significant simplifying assumptions. Finite element simulations, on the other hand, can be employed in conjunction with the experimentally measured profiles of the deformed particles for more precise and reliable constitutive modeling. Xie et al. (111) and Chen et al. (161) used post-impact measurements of the flattening ratio of particles in a range of impact velocities, from 50 to 950 m/s to optimize a bilinear Johnson-Cook material model. It was found that, for a successful prediction of the deformed profile, the strain rate sensitivity parameter in the Johnson-Cook equation should be increased by an order of magnitude, from $2 \times 10^{-3}$ at low strain rates to $2.9 \times 10^{-2}$ at strain rates higher than 600 s$^{-1}$.

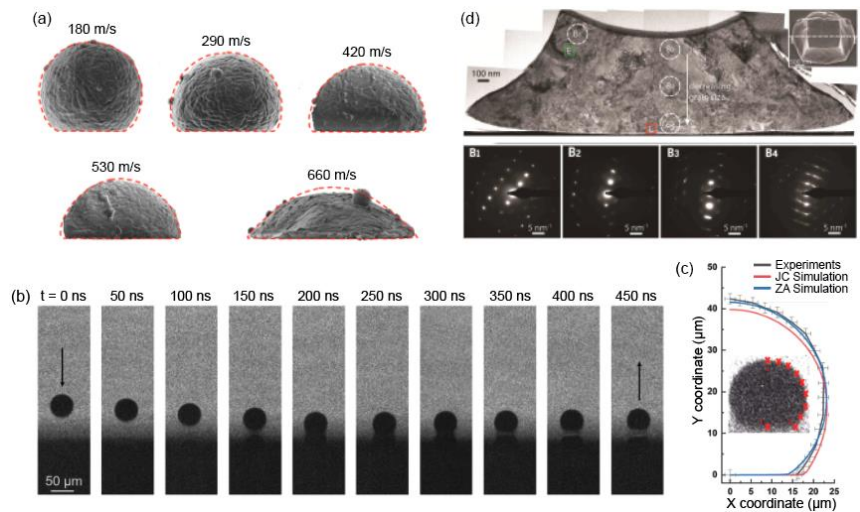

FIG 15. (a) Scanning electron microscope images of Al 6061 particles after impact onto sapphire showing the deformed shapes and flattening of the particles in a range of impact velocities. Red dashed lines indicate outlines of numerically simulated deformed particles (111). Reprinted with permission from Xie *et al.*, Sci Rep. **7**, 5073 (2017) under the terms of the Creative Commons Attribution 4.0 International License. (b) Real time observations of the impact-induced deformation of a 48-μm commercially pure titanium particle impacting a rigid alumina substrate at 190±5 m/s. The plot (c) next to the snapshots compares the experimentally measured deformed profile of the particle in (b) with the finite element simulations with optimized Johnson-Cook and Zerilli-Armstrong material models (162). Reprinted with permission from Wang *et al.*, J. Appl. Mech. **87**, 091007 (2020). Copyright 2020 ASME. (d) Cross-sectional bright field transmission electron microscope image of a silver microcube after a ~400 m/s impact along the ~[100] direction. Inset shows the impacted microcube from the top with a dashed line indicating the location of the cross-section. Selected area diffraction patterns confirm a variation from single-crystalline structure in the top region to the nanocrystalline structure at the bottom region (163). Reprinted with permission from Thevamaran *et al.*, Science **354**, 312 (2016). Copyright 2016 The American Association for the Advancement of Science

Wang et al. (162) also reported that the Johnson–Cook strain rate sensitivities measured by Klosky bar experiments lead to relatively inaccurate deformation when applied to modeling LIPIT impacts. They used in-situ observations of the deformation of commercially-pure Ti particles impacting a rigid alumina substrate, such as shown in the snapshots of **Fig. 15b**, for constitutive modeling. Two material models namely, Johnson–Cook and Zerilli–Armstrong, were implemented in iterative finite element simulations combined with an optimization scheme to generate matching deformed profiles with the experiments (see **Fig. 15c**). Close to fivefold increase in the Johnson–Cook strain rate sensitivity was found to be necessary to predict impact-deformation of Ti microparticles. The Zerilli–Armstrong model was found to have a better performance than the Johnson–Cook model which was attributed to the micromechanical origin of the former compared to the empirical nature of the latter. Wang et al. (162) also developed a scheme to determine the exact strain rates regime experienced during LIPIT impacts by computing the incremental plastic work and the strain rate for each volume element at every time interval. For a Ti particle impact at 190 m/s, it was found the entire plastic work is done at strain rates beyond $10^6$ s$^{-1}$. While the majority of the plastic deformation (almost 70%) occurs at the strain rate of ~$10^8$, strain rates exceeding $10^9$ s$^{-1}$ were also found in a smaller fraction (about 8.5%).

The higher strain rate sensitivities found at the LIPIT-associated strain rates (111, 161, 162) compared to Klosky bar calibrations can be mechanistically explained by the thermally-activated motion of dislocations as carriers of plasticity (164). The probability of dislocations to overcome barriers and obstacles in this mechanism follows an Arrhenius type relation and decreases at higher strain rates. As a result, higher flow stresses are required to maintain plastic deformation, giving rise strain rate strengthening. What is more, the interactions of a moving dislocation with phonons and electrons at higher strain rates produce non-negligible drag forces which in turn provide an additional strengthening mechanism.

In addition to constitutive modeling, Taylor-like impacts tests with LIPIT have been successfully used for studies of microstructural evolution and phase transformation under extreme conditions. Thevamaran et al. (163) have used near-defect-free single crystals of Ag in near-perfect cubic geometry as projectile and Si as rigid substrate. **Figure 15d** shows the cross-sectional transmission electron microscopy analysis of the microstructure of an Ag micro-cube after an impact at ~400 m/s along the [100] direction. A strong gradient in grain size can be observed along the height of the deformed microcube. The selective area diffraction patterns confirm a gradient microstructure, from a single crystal on the top to nanocrystalline at the bottom. The grain sizes on the top and at the bottom were measured to be ~500 nm and ~10 nm respectively, demonstrating a LIPIT-induced gradient that is at least an order of magnitude steeper than the typical gradients produced by the more conventional surface mechanical grinding and surface mechanical attrition. The extreme grain refinement at the bottom of the projectile was attributed to a shock wave-induced severe plastic deformation that together with the adiabatic heat generation resulted in dynamic recrystallization.

In another study, Thevamaran et al. (165) found a martensitic phase transformation from a face-centered-cubic (fcc) structure to a hexagonal-close-packed (hcp) structure during the impact of Ag microcubes at ~400 m/s. The phase transformation was found to be orientation dependent. While impact along the [100] direction resulted in the martensitic phase, impact along the [110] direction did not trigger any phase transformation. Molecular dynamic simulations showed that the impact-induced shock is the key driving force for generating avalanches of partial dislocations for the [100] impact which upon merging at the middle of the cube, initiate the phase transformation. The lack of abundant dislocation emissions and thus phase transformation in the [110] was attributed to a relatively lower impact-induced hydrostatic stress.

### C. SYNERGISTIC RESPONSE

The synergistic deformation of metallic microparticles impacting onto metallic substrates can give rise to the conditions necessary for solid-state bonding. Successive impact-induced bonding at the microscale is now routinely exploited for deposition of coatings, structural repair, and additive manufacturing via cold spray process. The same conditions can be induced in a LIPIT experiment enabling fundamental studies of impact-bonding that are otherwise challenging to conduct using cold spray nozzles. Our intention in this section is not to overview the progress made in in the past decades in the field of cold spray which have already been the subject of several books (166–170) as well as succinct and thorough reviews (171–178). Instead, our focus is to discuss how, in the past few years, LIPIT-based experiments have helped extend the understanding of the mechanical and materials phenomena relevant to cold spray.

A notable feature of the LIPIT is the ability to watch the interaction of a single particle and a substrate *in situ* and to develop understanding therefrom. Hassani et al. (116) resolved the moment of impact bonding in real time with micro-scale and nanosecond-level spaciotemporal resolution. The snapshots in **Fig. 16a** demonstrate the transition from rebound to bonding with increasing impact velocity for Al particles impacting an Al substrate. The particle accelerated to a velocity below the critical velocity for bonding rebounds despite a significant level of plastic flattening. The second particle accelerated to a velocity slightly above the critical bonding velocity adheres to the substrate with distinct unstable jetting and material ejection. These real time observations confirm the critical role of jetting in impact-bonding and are the point of departure to follow-up discussions on the shock-induced nature of jetting and a proportionality between the critical velocity and the spall strength (179–181). The large interfacial strain from jetting displaces and fractures the surface oxide layer and creates fresh metallic surfaces while the impact-induced pressure brings particle and substrate surfaces into intimate contact to form metallurgical bonding.

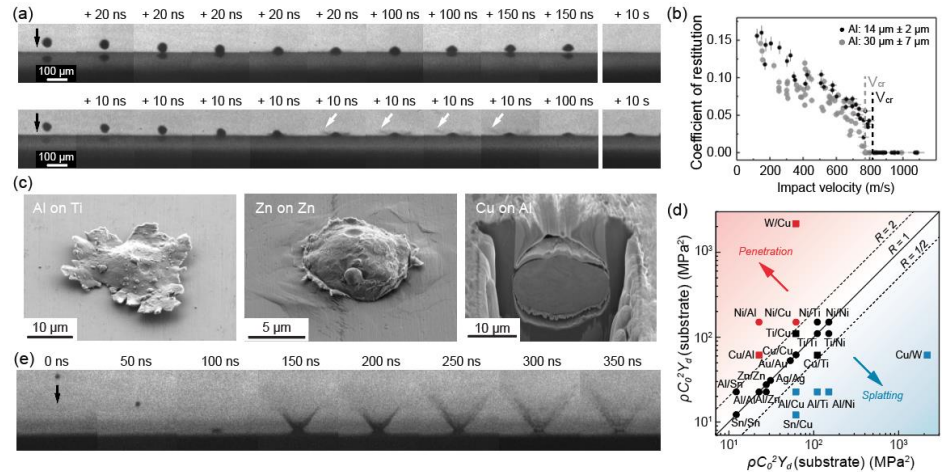

FIG 16. (a) In-situ observation of the bonding moment in microparticle impact. Multi-frame sequences showing 45-μm Al particle impacts on Al substrate at 605 m/s (top) and 805 m/s (bottom), respectively below and above critical velocity. The slower particle rebounds while the faster one bonds to the substrate. Material jetting is indicated with white arrows in the latter (182). (b) Coefficient of restitution for Al particles of two different sizes as a function of impact velocity (182). The drop of the coefficient of restitution to zero indicates the transition from rebound to bonding. Reprinted with permission from Hassani-Gangaraj *et al.*, Scr. Mater. **145**, 9 (2018). Copyright 2018 Elsevier. (c) Exemplar deformed geometries of bonded particles in the three regimes of splatting, co-deformation, and penetration (183). (d) Predictive map for impact-bonding regimes for similar and dissimilar materials with black data points indicating co-deformation, red data points penetration, and blue data points splatting. The diagonal line cutting the map into two corresponds to an impact ratio of unity around which co-deformation is the operative mode. Dotted lines correspond to ratios of ½ and 2 where transitions from co-deformation to splatting and penetration occurs respectively (183). Reprinted with permission from Hassani *et al.*, Acta Mater. **199**, 480 (2020). Copyright 2020 Elsevier. (e) Multi–frame sequences showing a 10-μm tin particle impacting a tin substrate at 1067 m/s and the resultant splashing and material loss in the erosion regime (184). Reprinted with permission from Hassani-Gangaraj *et al.*, Nat. Comm. **9**, 5077 (2018) under the terms of the Creative Commons Attribution 4.0 International License.

The separation between the two regimes of rebound and bonding can be precisely identified when plotting the coefficient of restitution as a function of impact velocity. **Fig. 16b** is an exemplar plot for Al particles impacting Al substrates (116). As impact velocity increases, more fraction of the kinetic energy of the incoming particles are dissipated by plasticity until a threshold where the coefficient of restitution drops to zero. Sun et al. attributed the deviation from the gradual decrease in the coefficient of restitution near the critical bonding velocity to the energy dissipated by jetting (185, 186). Critical velocities for twenty different combinations of particle and substrate materials were directly measured using the coefficient of restitution plots and reported in Ref. (183).

**Figure 16b** also shows that particle size can affect the critical velocity for bonding (116). Dowding et al. (187) conducted a focused study of the size effect on the critical velocity for Al and Ti particles impacting matched material substrates. It was found that that as particle size increases by a factor of ~4, Al and Ti critical velocities decrease by ~25%. The power law scaling between the critical velocity and the particle size measured via LIPIT experiments was comparable to the those measured via cold spray nozzle experiments (188). The mechanistic origin of the size effect is attributed to the fact that the higher kinetic energies carried by larger particles cause more local heating at the contact interface. Subsequently the local resistance of material to spallation decreases at higher temperatures. Dowding et al. (187) used the proposed proportionality between the critical velocity and the spall strength (179, 181) to predict the size effect and reported a reasonable agreement between the theory and the experimental measurements.

Jetting and material ejection were also observed during impact-bonding of gold particles on a gold substrate (189). With gold not having tendency to form a native oxide layer at the conditions of the experiment, this observation confirms that the spall fragments are ejected from the base metal. Nevertheless, the thickness of the native oxide layer can significantly affect the critical bonding velocity. Lienhard et al. (190) exposed Al powder particles to 300 °C for up to 240 min in dry air, or to room-temperature with humidity levels as high as 50% for 4 days and consequently conducted LIPIT experiments. No significant differences were found in terms of the critical bonding velocity which were attributed to the similar characteristics of the passivation layer (thickness, uniformity, crystallinity and composition) in different powder batches. An approximately 14% increase in the critical bonding velocity, on the other hand, was measured for powder particles exposed to 95% relative humidity for 4 days. The latter exposure resulted in a 2-3 nm increase in the thickness of the passivation layer as well as changes in the chemistry and structure of that layer. The increase in the critical bonding velocity was theoretically justified by considering the proportionality between the native velocity and the spall strength (179, 181).

The geometry of particle and substrate interfaces is an important factor in determining the bond strength and can significantly vary from one pair of particle and substrate materials to another. Hassani et al. (183) conducted LIPIT experiments on nine dissimilar particle/substrate material pairs in addition to their previous matched particle/substrate material impacts (116, 181, 184) and identified two limiting impact-bonding modes, namely, splatting and penetration. In both cases it is one material that accommodates a significant fraction of the impact-induced plasticity, i.e., the particle material in the splatting mode and the substrate material in the penetration mode. In between these two limits, lies a regime of co-deformation where a similar level of plastic deformation occurs in both particle and substrate materials upon impact-bonding. Show in **Fig. 16c** are typical examples of the bonded particles in the three regimes of splatting, co-deformation, and penetration based upon which a spectrum of impact modes was proposed (183). A dimensionless ratio—a function of density, bulk speed of sound, and dynamic yield strength of particle and substrate—was proposed to quantify the spectrum. **Figure 16d** shows a predictive map of impact-bonding modes based on the proposed ratio with the experimental observations also populated on the map. Co-deformation was found to occurs for the material pairs with ratios around unity, while splatting and penetration are dominant for ratios smaller than ½ and larger than 2 respectively.

The timescale of deformation in metallic microparticle impact (usually on the order of 10–100 ns) is shorter than the time needed for the plasticity-induced heat to be conducted away from the interface. The highly adiabatic nature of the deformation can, as a result, lead to localized interfacial melting at high impact velocities which in turn can interfere with solid-state bonding. In fact, the localized melting at substrate surfaces in Al particle impact on Zn and Sn was shown to hinder bonding (191). The resolidification of the molten layer takes orders of magnitude longer than the time that the particle resides on the surface of the substrate. In other words, when localize melting occurs the particle is mostly in contact with a liquid interface whose low mechanical strength is easily overcome by a rapidly rebounding particle.

A major fraction of the impact-induced plastic work dissipates as heat. If the heat is high enough to melt materials beyond the interfacial regions, then impact can cause material loss or erosion. The phenomenon, resolved by Hassani et al. (184), is shown in real time in the snapshots of **Fig. 16e** where a Sn particle impacts a Sn substrate at ~1 km/s. A splash with a cloud of ejecta is observed leading to a loss of material on the order of ~100 $\mu m^3$. While the critical bonding velocity represents the transition from the rebound to the bonding regime, the threshold velocity to induce material loss, the so-called erosion velocity, marks a second transition from the bonding to the erosion regime. The critical and the erosion velocities together define the lower and the upper bounds to the deposition window, i.e., the range of impact velocities that lead to successful material build up in cold spray. Erosion maps developed based on the mechanistic origin of material loss can be used to calculate the erosion velocity for given particle/substrate materials and a given particle size (184).

## IV. PERSPECTIVES

There are emerging opportunities in materials science that involve utilizing microscopic high-speed impact of a microparticle with precisely-defined collision parameters and geometries (projectile material, size, shape, temperature, impact velocity, impact angle), as a highly localized and quantifiable mechanical stimulus of high strain rates becomes available by LIPIT. While the momentum associated with the microscopic single impact is enough to produce HSR plastic deformation of a specimen, the resultant deformation volume (~1000 $\mu m^3$) is still small enough to be considered as a nondestructive characterization that can be combined with most high-resolution characterization techniques. By virtue of this unique aspect of LIPIT, HSR elasto-plastic and visco-plastic behaviors of various materials including metals, polymers, and nanomaterials have been investigated, which would be difficult by other means. In contrast to the well-quantified kinetic parameters in LIPIT, the measurement of materials state parameters such as pressure, temperature, stress and strain remain a challenge. Therefore, we foresee two major avenues for development of HSR studies using microparticle impacts.

First, most techniques presented in section II rely on high-speed imaging for both launch characterization (particle size, speed, distribution, etc.) and sometime impact response diagnostic (real-time imaging of penetration, target damage). While post-mortem examinations inform greatly about the impact history, real-time measurements represent the ultimate goal. Currently, laser-flyer techniques, compared to other methods, can leverage their optical nature to implement the most advanced synchronized optical diagnostics. These advanced diagnostics, such as optical pyrometry, real-time Raman spectroscopy, or femtosecond X-ray scattering measurements (192) have yet to be applied for LIPIT studies.

Second, remaining challenges may be difficult to overcome solely with the advance of LIPIT, due to the strong locality of the transient fundamental quantities. The next major step in the understanding of the fundamental HSR materials properties is expected to be made through the synergistic combination of LIPIT and corresponding numerical modeling, where realistic HSR material parameters will be attained from the real-time LIPIT data including projectile dynamics and post mortem characterizations. For example, the ability to study plastic deformation of projectiles both *in-situ* and *post-mortem* has provided LIPIT with unique potentials for Taylor-like impact experiments and constitutive models development and refinement at ultra-high strain rates.

Further, as illustrated in **Fig. 9**, no method exists for single microparticle acceleration to velocities of the order of 10 km/s or higher. Reaching such velocities will open the door to more systematic studies in aeronautics and space science. The LIPIT apparatus is not limited by laser energies but rather by the resistance of launch pad to plasma expansion and by the resistance of the particle to acceleration. Through optimization of the launch pad design, using tougher polymers for example, higher velocities might be reachable, hence positioning LIPIT as a competitor in terms of velocity performance to plasma drag acceleration techniques. With higher velocities, the LIPIT technique could find use in the study of materials and structures for hypersonic applications. New materials, such as nano-structured refractory alloys and ceramic composites, could be screened for performance in hypersonic environments and velocities. In addition, slightly larger samples could be impacted with particles launched to hypervelocity by the LIPIT technique and atmospheric erosion and damage could be characterized.

The range of materials will continue to expand as novel nano-materials and architectures are designed and new physics are yet to be explored under high rate dynamic conditions. For instance, LIPIT has proven to be an effective platform for fundamental studies of the processing science of cold spray. The strength of LIPIT in this context stems from the fact that it decouples the multivariable influence in a complex deposition problem and enables studies of each variable at a time. While the effects of particle size, native oxide layer, alloying elements, mismatched material pairs on impact-bonding have been isolated and studied through LIPIT, the role of several other parameters such as impact angle, particle shape, particle/substrate temperature, and surface roughness is yet to be studied. The research efforts on this track have so far mainly focused on the impact behavior of metallic particles and substrates. With the rapid extension of kinetic deposition beyond metals and alloys, understanding impact and bonding in cases where particle and substrate belong to different classes of materials with substantially different properties would be highly desirable. Examples include metallic particles impact on polymers or ceramic particles impact on metals. During material buildup, impacts of particles onto particles are by far more probable than impacts of particles onto the substrate. Therefore, studies of particle-particle interactions would be also desirable.

## SUPPLEMENTARY MATERIAL

The data points used to plot **Fig. 9** are available as supplementary material online.

## CONFLICT OF INTEREST STATEMENT

The authors declare no conflict of interest.

## ACKNOWLEDGMENTS

DV, SEK, and KAN acknowledge funding support from the U. S. Army Research Office through the Institute for Soldier Nanotechnologies, under Cooperative Agreement Number W911NF-18-2-0048. JHL acknowledges funding support from the U.S. Department of Defense under Cooperative Agreement No. HQ0034-15-2-0007. MH acknowledges funding support from the US Army Research Laboratory through a cooperative research agreement W911NF1920329.

## DATA AVAILABILITY

The data that supports the findings of this study are available within the article and its supplementary material.

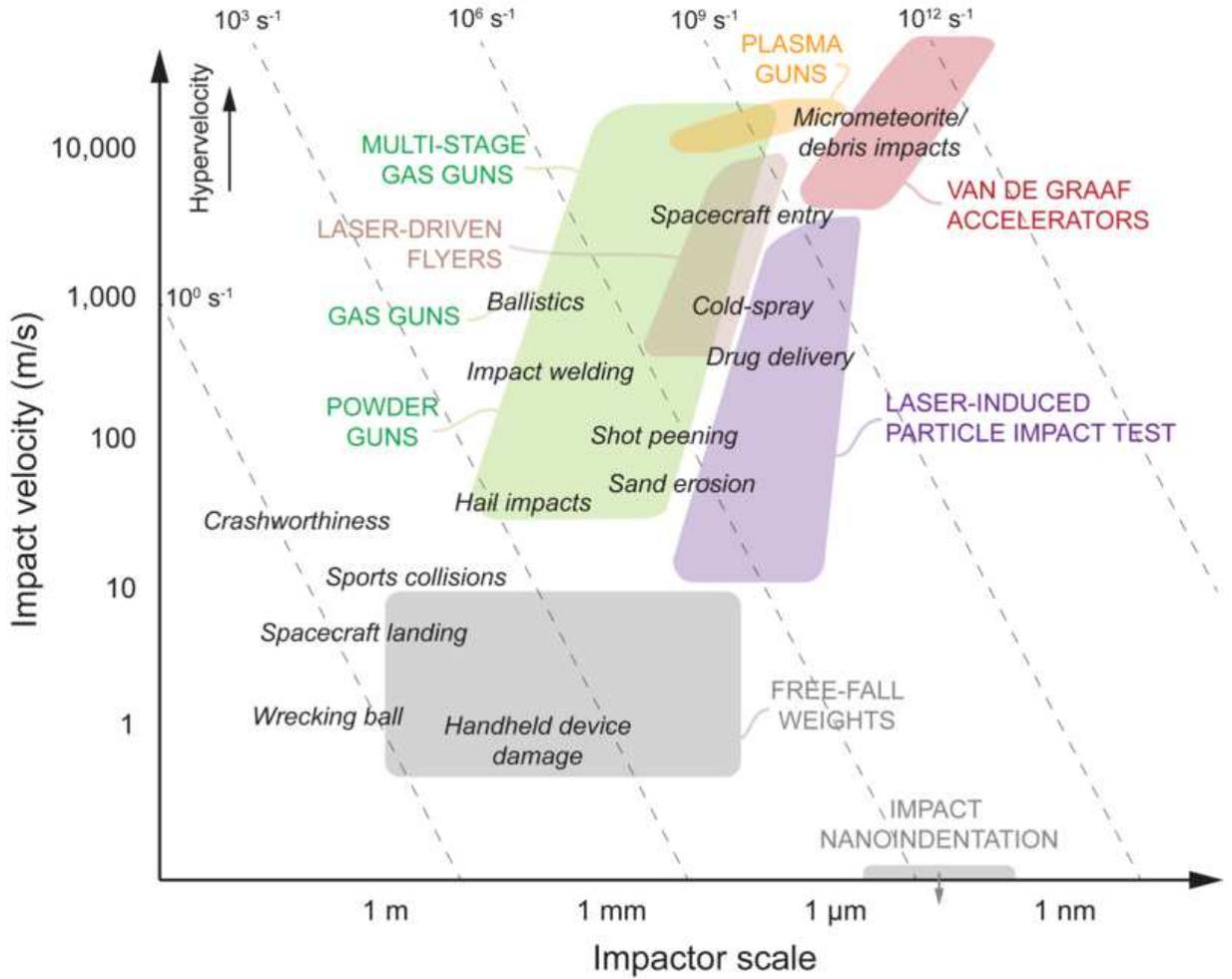

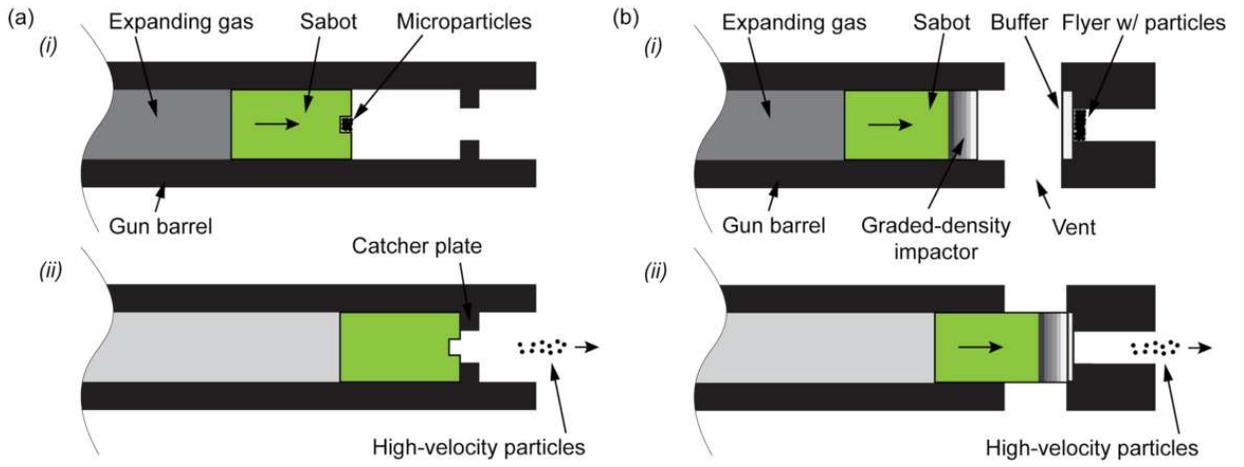

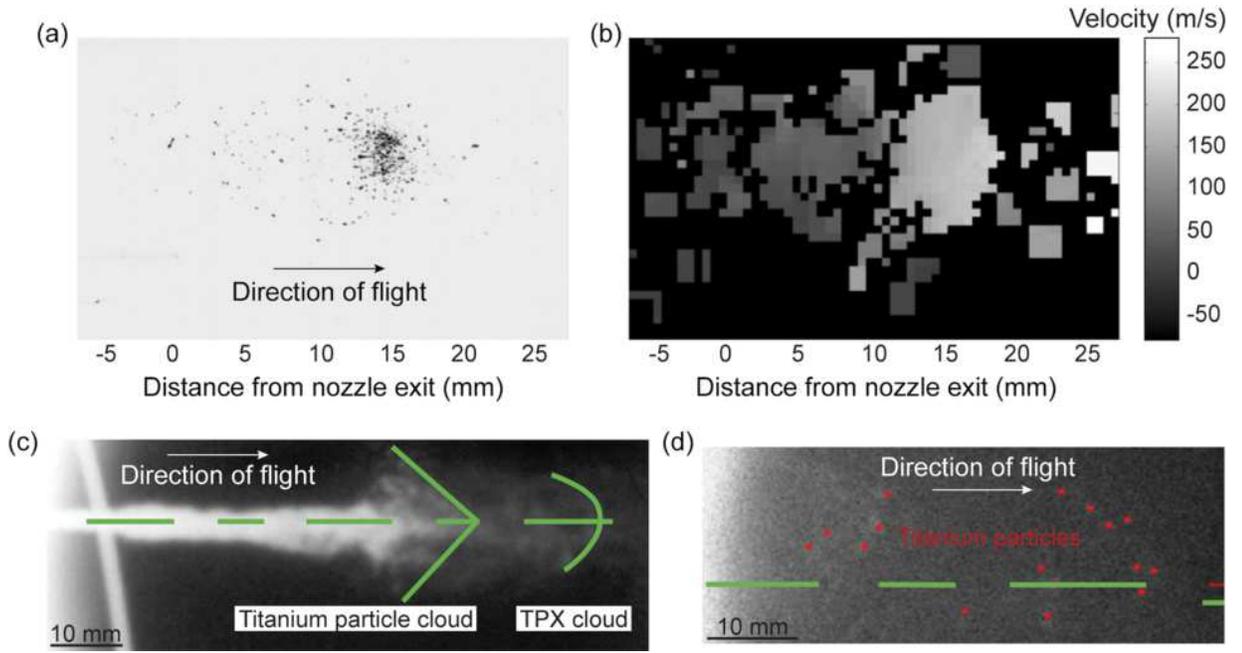

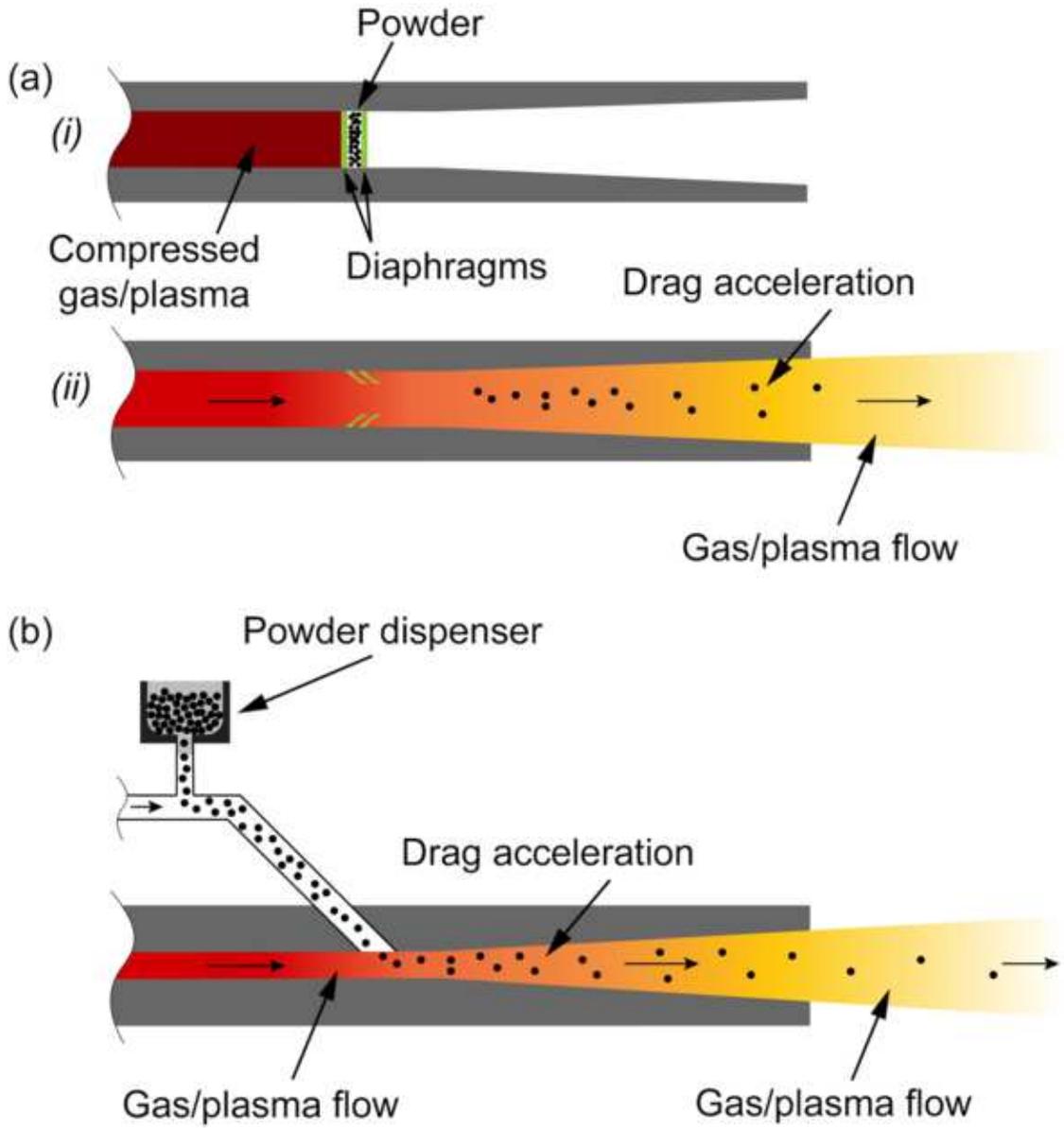

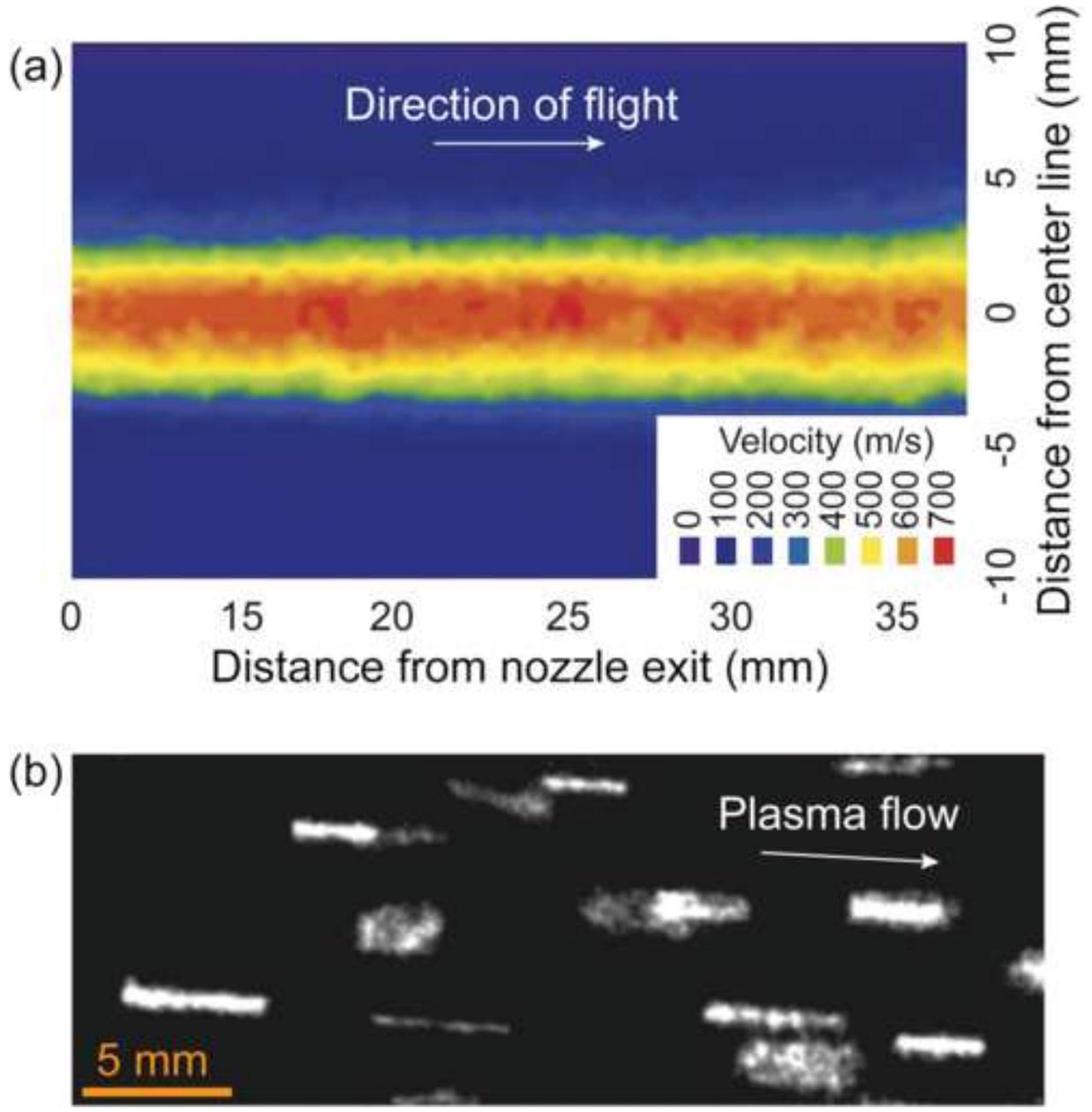

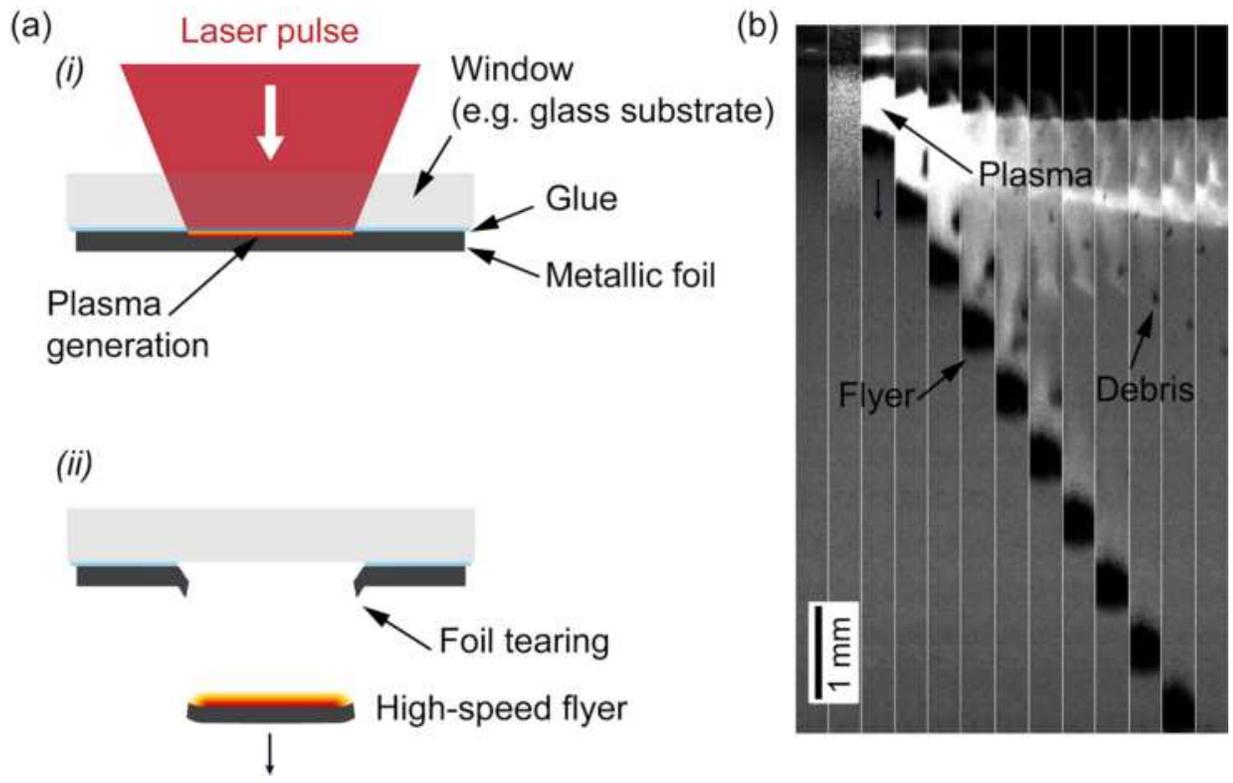

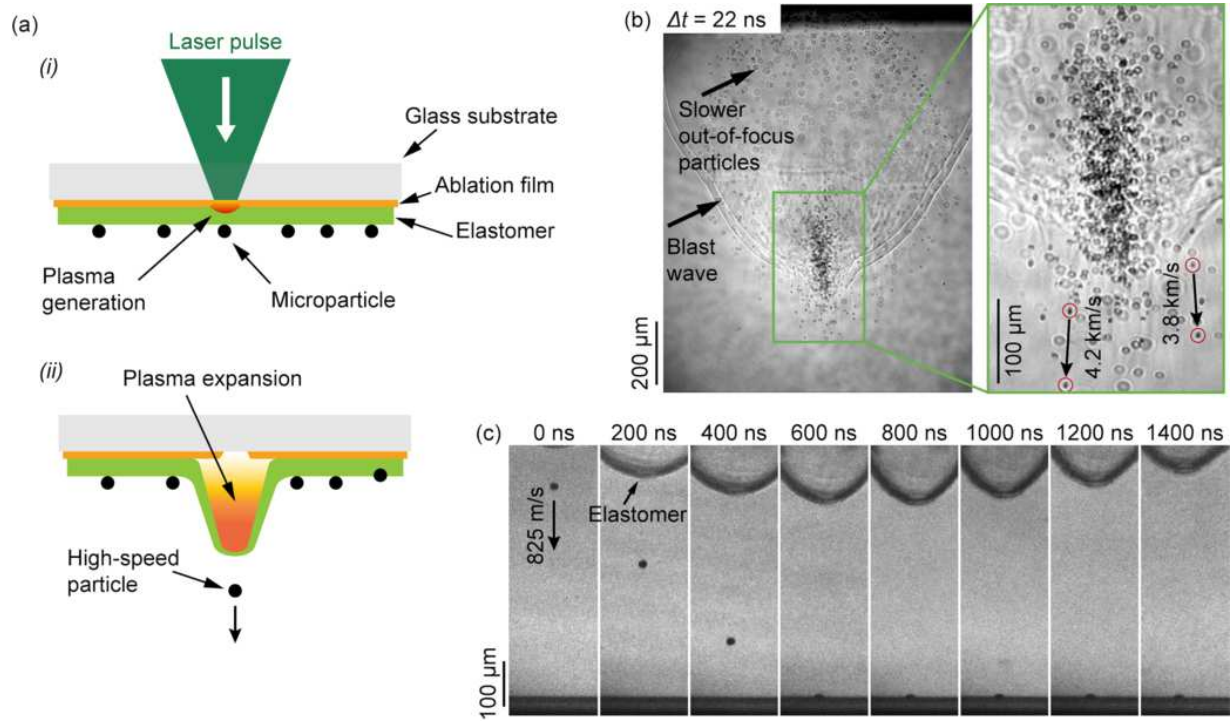

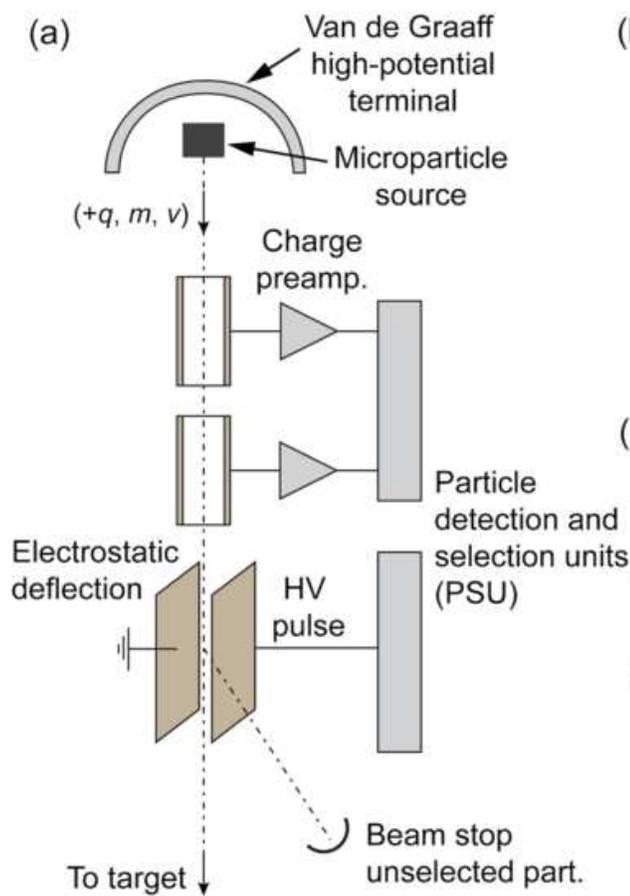

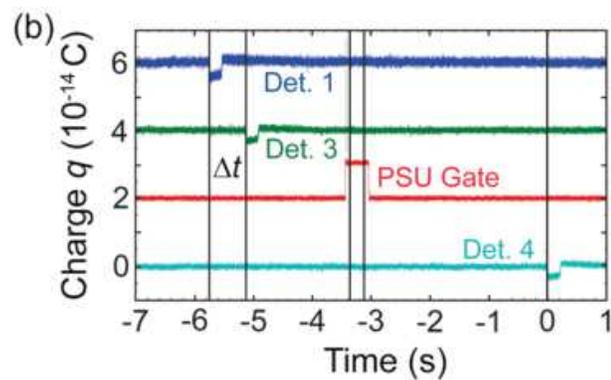

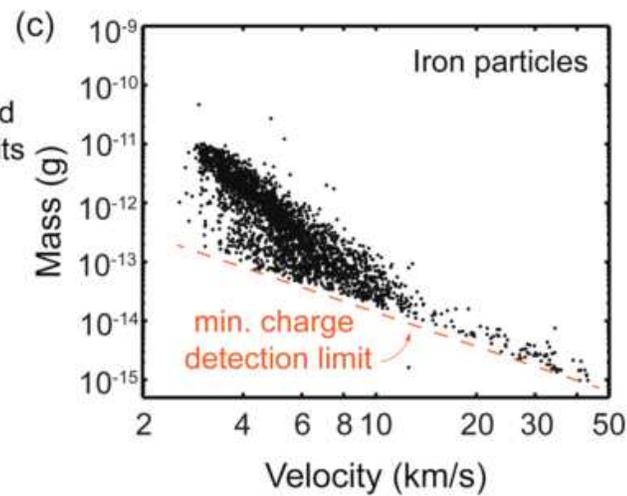

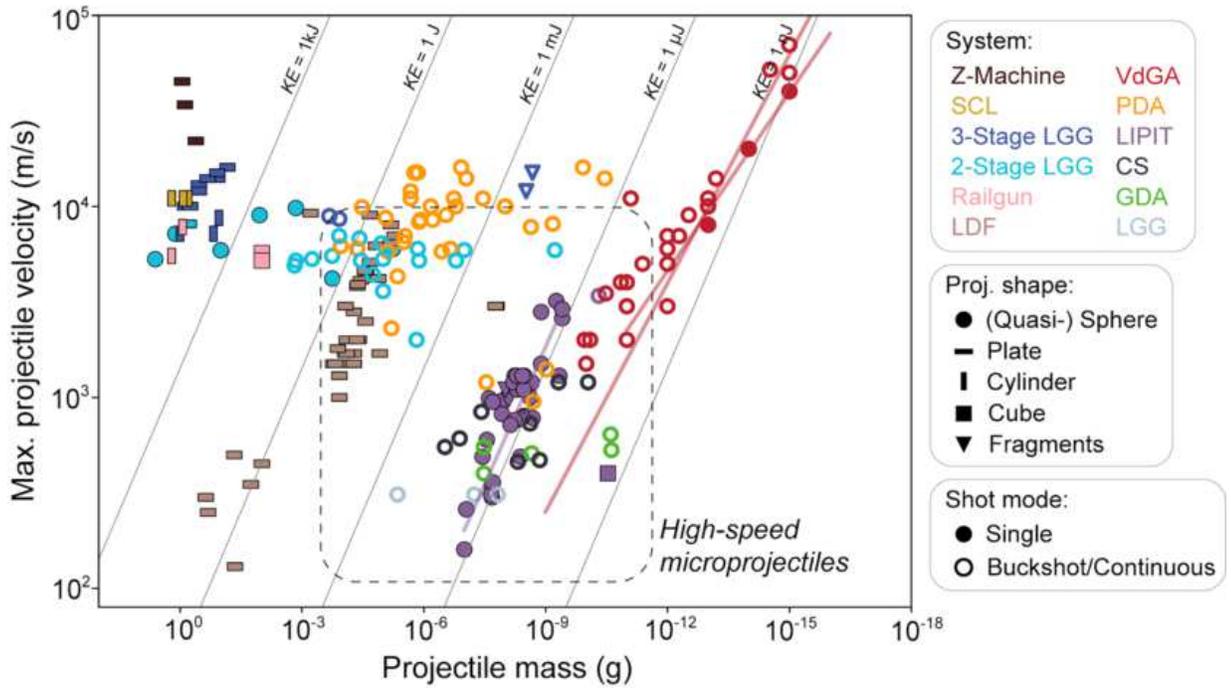

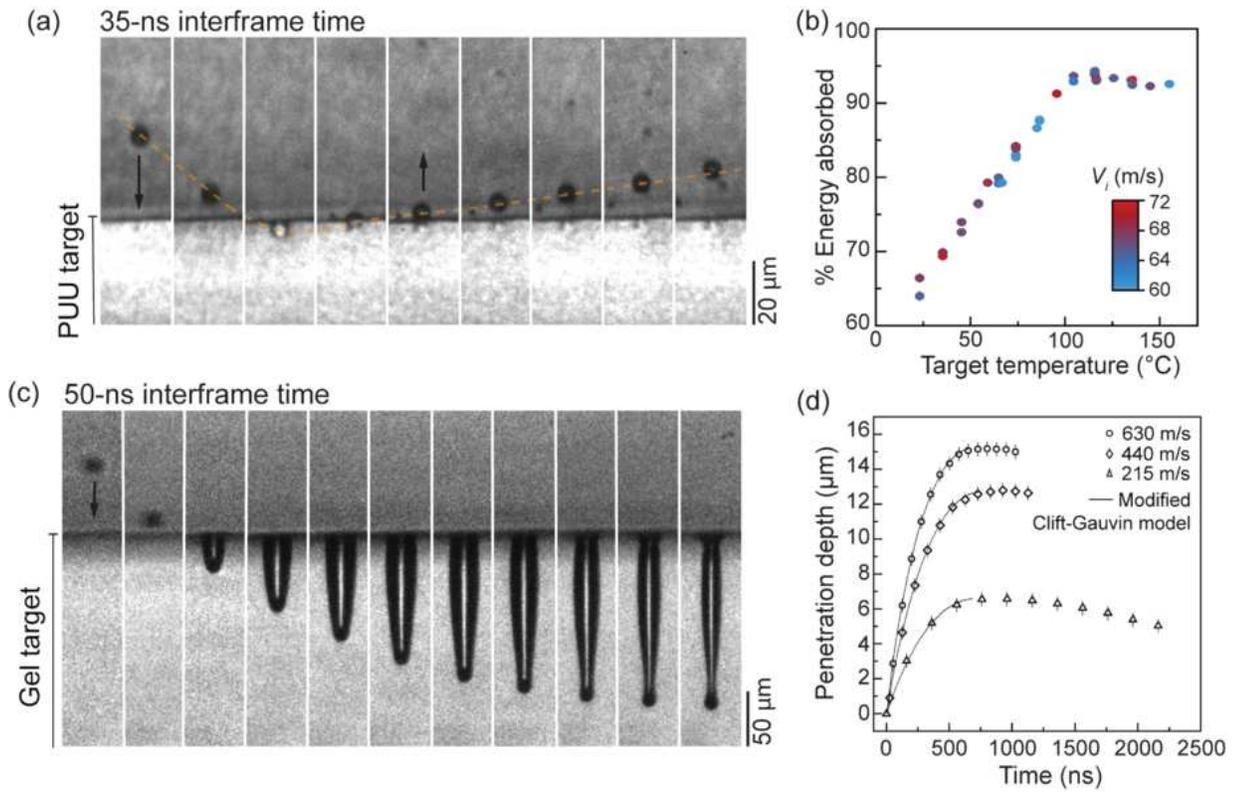

AIP
Publishing

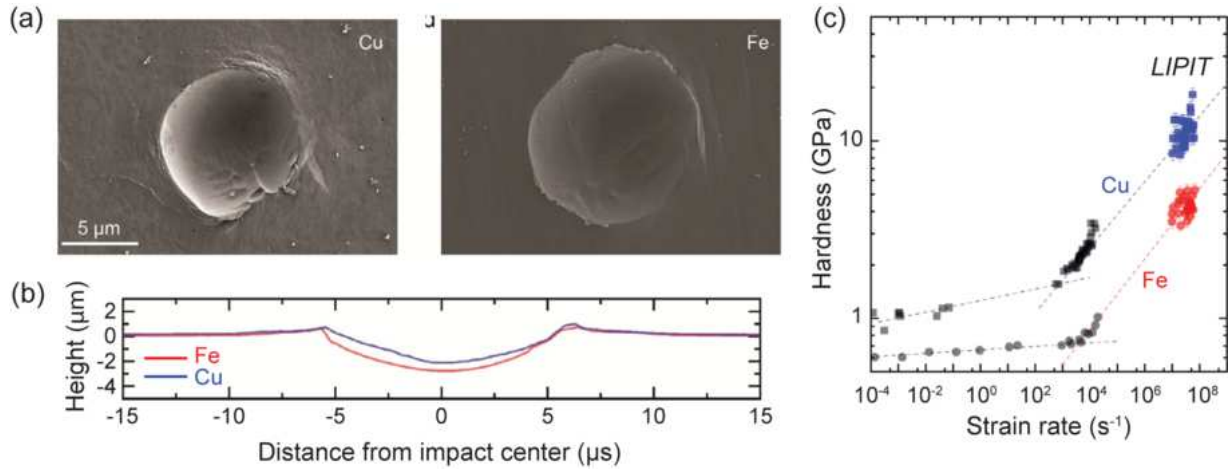

(a)  Primarily localized deformation

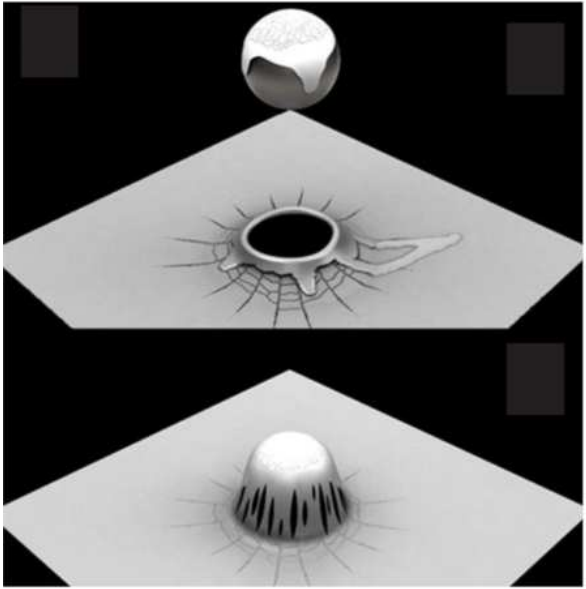

(b)  Primarily delocalized deformation

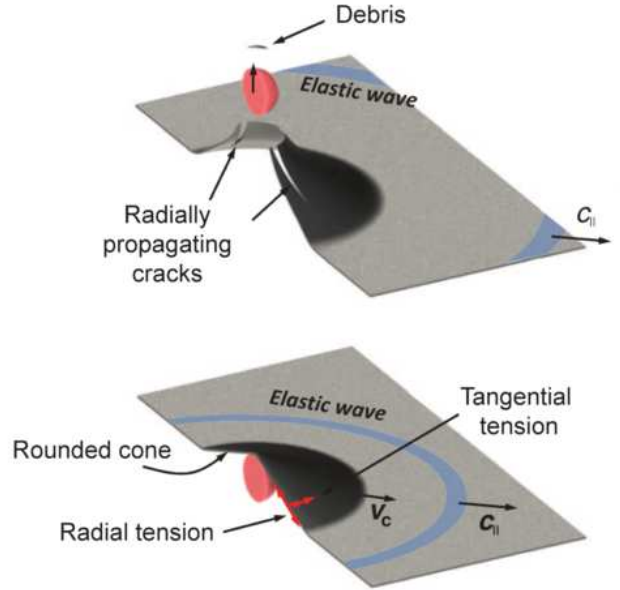

(a) Molecular weight effect (polystyrene)

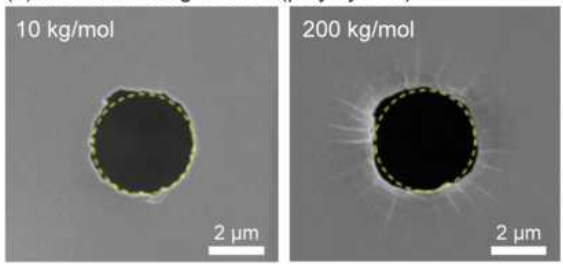

(b) Entanglement effect (polystyrene, polycarbonate)

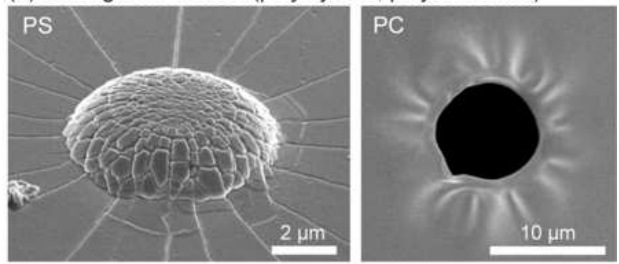

(c) Orientation effect (ordered block copolymer)

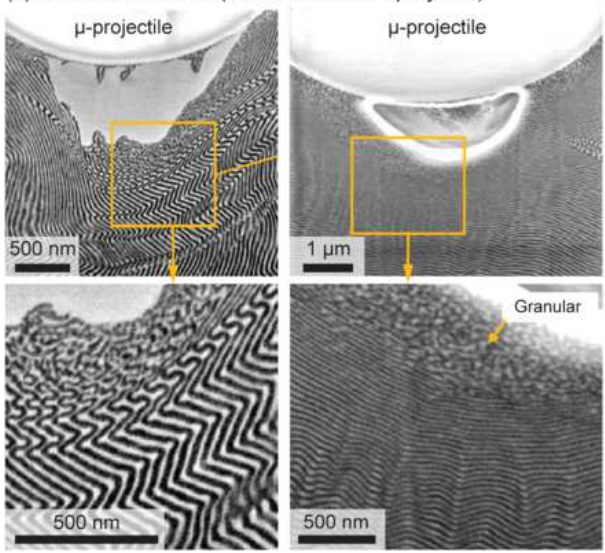

(d) Delamination effect (multi-layer graphene)

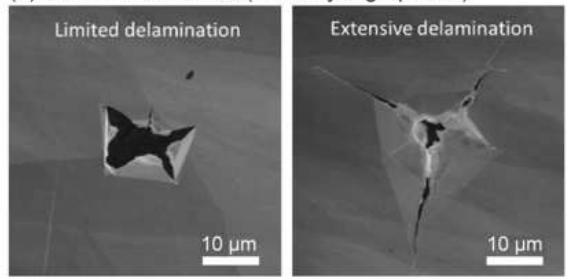

(e) Phase percolation effect (graphene-oxide/silk)

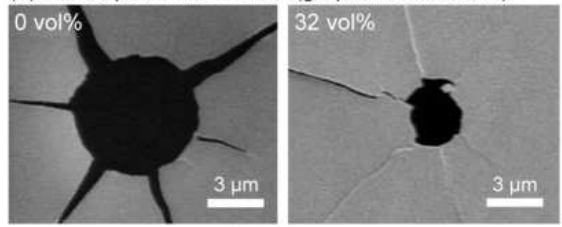

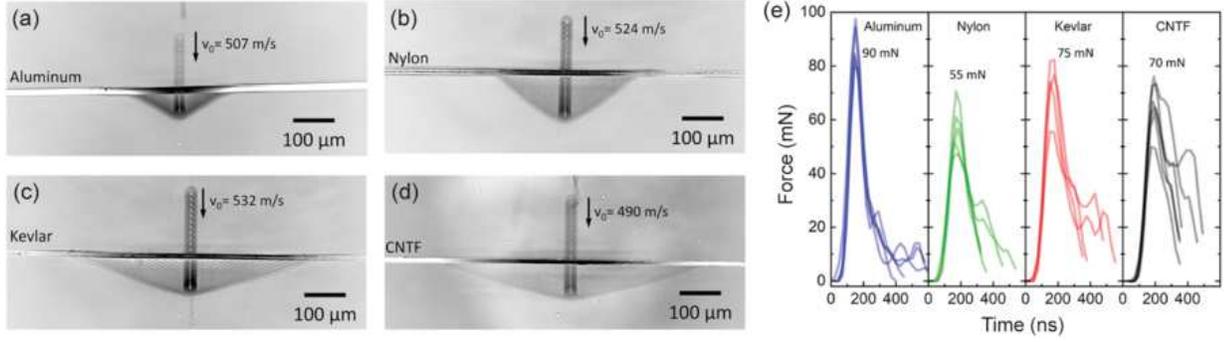

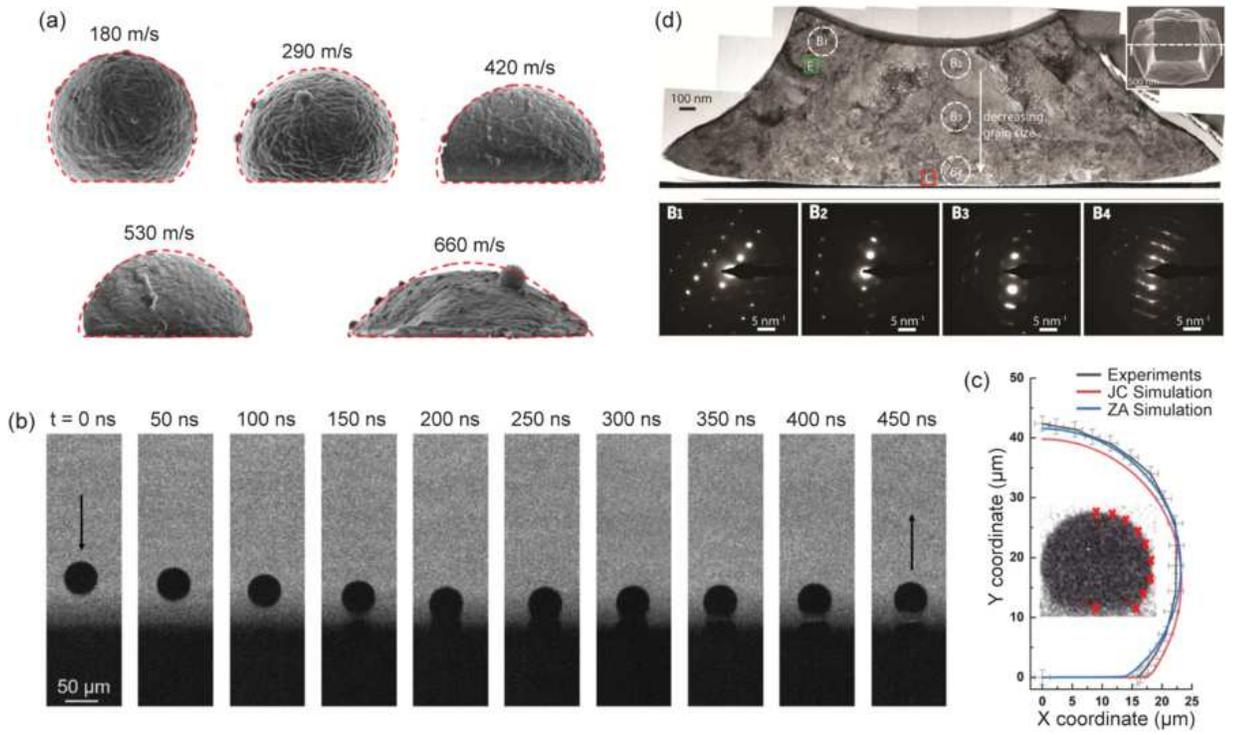

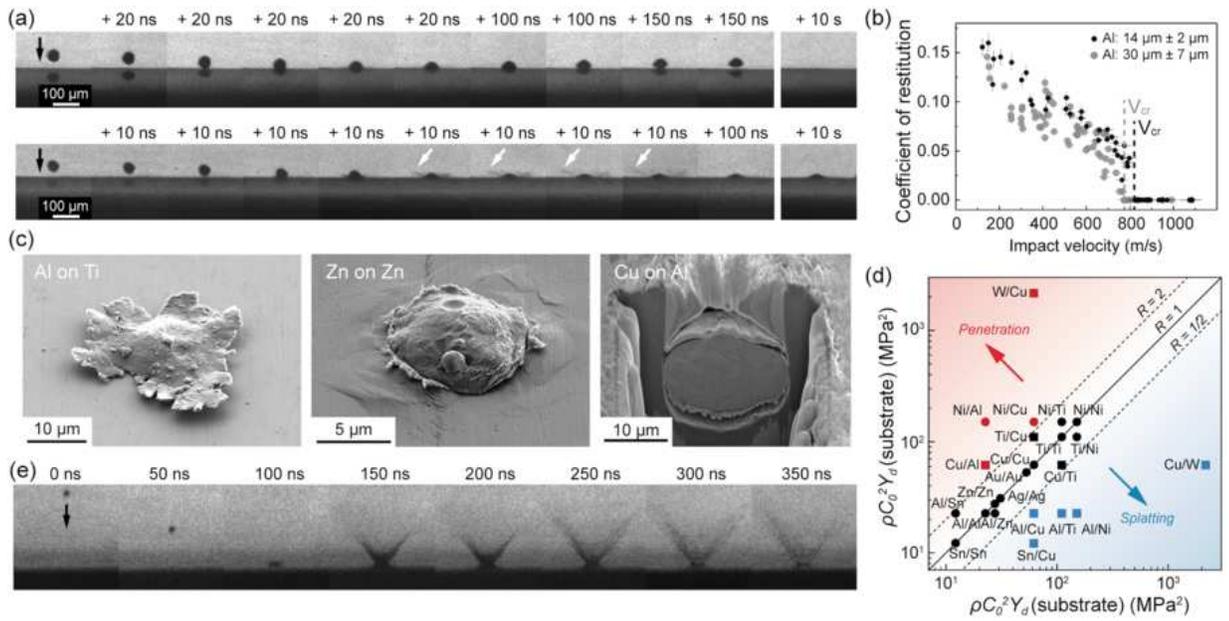